\documentclass[twocolumn,secnumarabic,amssymb, nobibnotes, aps, prd]{revtex4-2}
\usepackage{graphicx}
\usepackage{subfigure}

\newcommand{\env}[1]{\texttt{#1}}
\usepackage{bm}
\usepackage{natbib}
\usepackage[usenames,dvipsnames]{xcolor}
\usepackage[normalem]{ulem}
\usepackage{physics}
\usepackage[mathlines]{lineno}

\DeclareMathAlphabet\mathbfcal{OMS}{cmsy}{b}{n}

\begin{document}

\title{Chirality and Rashba-related Effects in the Spin Texture of a Two-Dimensional Centro-Symmetric Ferromagnet: the Case of CrI$_{3}$ Bilayer}%

\author{Sukanya Ghosh}
 \altaffiliation[Presently address : ]{Materials Theroy Division, Department of Physics and Astronomy, Uppsala University, 75120 Uppsala, Sweden}
\author{Nata{\v s}a Stoji{\'c}}%
\author{Nadia Binggeli}
\affiliation{%
 Abdus Salam International Centre for Theoretical Physics, Trieste, Italy
}%

\email[email : ]{sukanya.ghosh@physics.uu.se}
\date{December 2018}%
\maketitle







\section{abstract}


The newly discovered two-dimensional (2D) magnetic semiconductors such as  CrI$_3$ have triggered a surge of interest stemming from their exotic spin-dependent properties and potential applications in spintronics and magneto-optoelectronics. Using first-principle density-functional-theory calculations, we investigate the properties of the spin-polarization texture in momentum space in the prototype 2D centrosymmetric ferromagnetic (FM) bilayer of CrI$_3$, with perpendicular magnetization. The FM bilayer displays a rich in-plane spin texture in its highest valence bands. We show the existence of two distinct spin canting effects, which result from the coupling of FM order with structural chirality and electric polarization, in establishing the in-plane spin texture. The first effect is generated by the chirality of the layer stacking and the spin-orbit-polarized nature of the valence states, and yields the same canting on both layers. The second effect is a Rashba-related effect, which in a centrosymmetric ferromagnet induces in a single electronic state two opposite spin-canted components on the two layers. 
Furthermore, using the FM bilayer as an example, we provide some general rules for centrosymmetric systems imposed by magnetic-space-group symmetries on spin-polarization vectors, serving as guidelines for the shape of the spin texture in the Brillouin zone of such magnetic crystals. Finally, we show that the above effects can be effectively used to manipulate the spin texture via compressive strain, which induces in the FM bilayer valence-band-edge states with canted spins.





\renewcommand \thesection{\Roman{section}}
\renewcommand \thesubsection{\Roman{section}\alph{subsection}}

\section{Introduction}
\label{sec:intro}
%

Spin-orbit-coupled crystalline materials offer a fruitful ground for the  observation of a variety of spin-related phenomena and opportunities for controlling spin-dependent properties by non-magnetic means via spin-orbit coupling (SOC).\cite{Manchon_2015, Soumyanara_2016,Manchon_2019}  One of such properties of relevance for spintronics is the spin texture of electronic bands in momentum space, as it determines how spin-polarized currents may be manipulated for spintronic devices. 

Spin polarization in nonmagnetic materials has been traditionally ascribed to SOC in the presence of broken global inversion symmetry. The properties of the resulting non-trivial spin textures in momentum space have been widely studied for various classes of three-dimensional (3D) and two-dimensional (2D) non-magnetic crystalline systems lacking inversion symmetry.\cite{Schliemann_2017,Tsymbal_NatCom_2018,Tsymbal_HfO2_PRB_2017,perovskite_Xie_2018,Sakano_PRL_2020,Zhonghao_2020,Zollner_PRR,Huang_Scireport2020, ZhaNakArr20,Stroppa_2021,Zunger_2021} This includes, in particular, crystal surfaces and interfaces, where SOC effects are enhanced by reduced dimensionality.\cite{Soumyanara_2016} Furthermore, for nonmagnetic materials, general rules have been established, which provide simple symmetry-based guidelines on the shape of spin textures in the Brillouin zone  (BZ).\cite{Zunger_2021} 
In recent years, the discovery of hidden spin polarization\cite{Zunger_NatPhys_2014,YuaLiuZha19} and spin-layer locking\cite{YaoWanHua17} in inversion-symmetric nonmagnetic semiconductors and semimetals has steered research interest towards centrosymmetric materials. This recent development has also shown that specific local-site inversion asymmetry combined with SOC could produce remarkable spin-polarization effects in the spin textures of nonmagnetic centrosymmetric crystals.\cite{Zunger_NatPhys_2014,YuaLiuZha19,YaoWanHua17} 

In the past few years, an exciting new class of materials, the 2D magnetic materials, has been discovered.\cite{Gong_NatPhys_2017,HuangNature2017}  These materials are attracting enormous attention due to their peculiar spin-related phenomena and vast potential for magneto-optics, magneto-electronics, and spintronic devices.\cite{Burch_Nat2018,GibKop19,Soriano_20}  
This new class of materials belongs to the fast growing family of 2D van der Waals (vdW) materials, 
whose heterostructures hold great promises for nanoscale devices. 
Many of the 2D magnetic materials are semiconductors with centrosymmetric atomic geometries, e.g.,  CrI$_{3}$, CrBr$_{3}$, CrCl$_3$, Cr$_{2}$Ge$_{2}$Te$_{6}$, and show ferromagnetism or layer-dependent ferromagnetism.\cite{Gong_NatPhys_2017,HuangNature2017,Chen_2019,Cai_2019} Importantly, the magnetism in such ultrathin magnetic materials can be tailored by external perturbations such as electric field,\cite{HuaClaKle18,Jiang_2018} pressure,\cite{Song_2019,Li_2020} doping,\cite{Jiang_Nat_Nanotech_2018,Wang2018_NatNanotech,Verzhbitskiy_2020} and stacking engineering.\cite{Chen_2019,Song_2019,Li_2020} 

The emergence of the 2D magnetic semiconductors is opening the door to the design and fabrication of novel vdW spintronic, \cite{Song_Sci_2018,Cardosa_2018,Klein_Sci_2018,Wang_nat_commun_2018,Dolui_2020}  magneto-optical\cite{Stroppa_2020} and magneto-(opto)electronic\cite{Zhang_2019,Cheng2021}  devices.
In view of the incorporation of 2D magnetic materials in vdW devices it is important to know and understand the properties of their band-spin texture and the mechanisms that control them ---in order also to manipulate them. General symmetry-based rules serving as guidelines on the shape of spin textures in the BZ of magnetic materials would also be helpful. 
So far, however, the band-spin textures in momentum space of magnetic materials with inherent SOC, and in particular of ferromagnets, have remained largely unexplored. Only recently have a few investigations started to examine magnetic-order effects on the spin texture of some  antiferromagnets,\cite{Barone_PRB_2019,Stroppa_npjcomput_2020,Zunger_2020,AFM_SG} while other studies have focused on surface/interface states and indicated Rashba effects and helical spin texture at ferromagnet surfaces/interfaces.\cite{Blugel_2005,Tsymbal_2016,Margalit_2020}

Here we explore the properties of the band-spin texture in momentum space for a ferromagnetic (FM) vdW bilayer of CrI$_3$, based on first-principles calculations. 
This system exemplifies the effects of chirality and electric polarization combined with FM order in establishing the spin-texture features of a relativistic centrosymmetric magnet.  
CrI$_3$ is a particularly interesting example of 2D vdW magnetic semiconductors, with perpendicular magnetization easy axis. The CrI$_{3}$ monolayer is FM, whereas the CrI$_{3}$ bilayer in its most common experimental form is a layered antiferromagnet,\cite{SunYiSon19,HuaClaKle18} with a weak interlayer exchange coupling. 
This makes possible switching experimentally from the AFM to the FM phase with low magnetic field,\cite{HuangNature2017} and most remarkably also   
by electrical means, such as electrostatic doping and electric field.\cite{HuaClaKle18,Jiang_Nat_Nanotech_2018} Change in the layer stacking was also shown to change the magnetic coupling from AFM to FM.\cite{Li_2020,Sivadas_NanoLett,Jiang_PRB_2019,SORIANO_ssc} 
CrI$_3$ possesses rather strong Cr magnetic moments oriented perpendicular to the layers and a hefty Iodine SOC.\cite{SG_2021} Earlier studies of the bilayer and monolayer have shown that SOC has a major impact on the highest valence states, while the lowest conduction bands remain  essentially unaffected by relativistic effects.\cite{SG_2021,Lado_2017} 

In this work we examine the spin texture for the highest valence states of the CrI$_{3}$ FM bilayer, and compare to the FM monolayer and AFM bilayer.\cite{AFM_SG} In contrast to the latter cases, the FM bilayer displays a rich in-plane spin texture for the highest valence band. It includes  alternating radial-spin features and vortices which are induced around the BZ center by the interlayer interaction. We identify the mechanisms leading to such features and their specific chiral origins. We also investigate the canting of the spin-polarization vectors on each of the CrI$_3$ monolayers, and disclose an interesting Rasbha-related frustration effect present in such centrosymmetric ferromagnets. 
Using the FM bilayer as an example, we determine general rules imposed on spin textures by magnetic-group symmetries. We show that symmetry operators that combine time reversal with rotations and reflections in magnetic groups of centrosymmetric systems give rise to rules for the spin texture in the BZ which are reversed with respect to those established for non-magnetic groups ---i.e., they impose that spin-polarization vectors are parallel to mirror planes and perpendicular to rotation axes for such operators. Finally, we show for the CrI$_3$ FM bilayer  that vertical compressive strain can be used to effectively manipulate the spin texture and induce valence-band-edge states with canted spin polarization. 

The remainder of this paper is organized as follows: in the next section the main structural and magnetic properties of the bilayer and monolayer CrI$_3$ are described, which is followed by the account of  methods and parameters of calculations.  The results part of the paper starts with the presentation of the spin texture of the highest valence state on the FM CrI$_3$-bilayer in $k$--space in Section~\ref{sec:pristine}, while the layer dependence of the spin--polarization density is presented  in Section~\ref{sec:profile}.  The effect of the FM symmetries on the spin texture is presented in the subsequent  section, and the two mechanisms responsible for the formation of the bilayer spin texture, namely the interactions through the local electric field (Rasbha effect) and through the interlayer chiral potentials,  are explained in Section~\ref{sec:mechanism}. After that, in Section~\ref{sec:stress}, we show how the spin texture can be manipulated and the summary of the main results is given in Conclusions. The appendix contains information on the monolayer spin texture and band structure, the spin texture of the second valence state of the bilayer and explains further the calculation of the Rashba and chiral-interaction terms.

\section{System and Symmetries: Bilayer versus Monolayer}
\label{sec:system}

\begin{figure*} [t]
\includegraphics[width=0.8\textwidth]{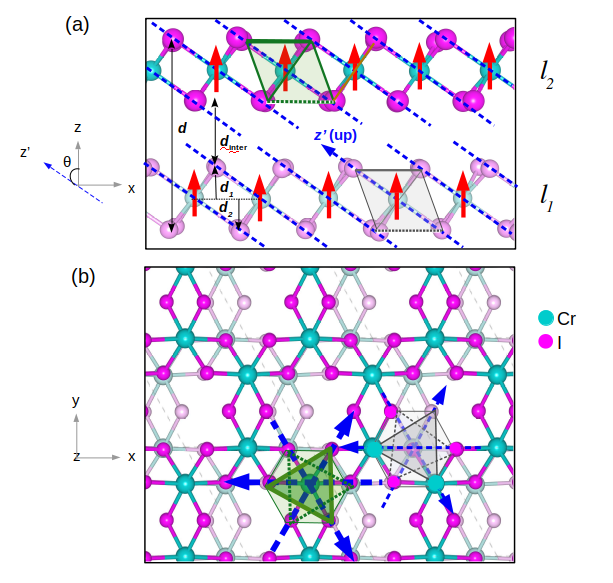}
\caption{\label{structure} Side view (a) and top view (b) of the CrI$_{3}$ FM bilayer with monoclinic layer stacking.  The top (bottom) CrI$_{3}$ layer is indicated with bright (light) colors. The red arrows show the directions of the Cr spin moments. 
Each Cr atom is surrounded by six nearest neighbour I atoms, forming an  octahedron, as indicated in green (grey) for the top (bottom) CrI$_{3}$ layer. The upper (lower) facet of the octahedron is shown by the solid (dotted) triangle.  Dashed blue lines in (a) show the body diagonals of the iodine octahedra enclosed in planes parallel to the $xz$ plane; these diagonals are aligned along the $z'$ axis. The blue arrows in (b) show the projections of the directions of the body diagonals of octahedra oriented from bottom to top I-layer atoms.  
The thin black arrow  in (a) shows the distance $d_{1}$ ($d_{2}$) in a CrI$_{3}$ monolayer  between the Cr atomic plane and the I innermost (outermost) plane of the bilayer. The distance between the outermost (innermost) I planes of the bilayer $d$ ($d_{int}$) is also shown.}
\end{figure*}

In this study we focus mainly on the CrI$_3$ FM bilayer with monoclinic layer stacking, displayed in Fig.~\ref{structure}. 
In each of the CrI$_3$ monolayers, shown in this figure, the plane of Cr$^{3+}$ ions is sandwiched between two planes of I$^{-}$ ions, and the  Cr$^{3+}$ ions form a honeycomb lattice. Each Cr$^{3+}$ ion has six nearest-neighbor I$^{-}$ ions (three from the upper I plane and three from the lower I plane) forming a network of edge-sharing iodine octahedra. The spin magnetic moment per Cr is $3 \mu_B$.

The CrI$_{3}$ bilayer is observed experimentally in two different structural phases which differ in their van-der-Waals CrI$_{3}$-layer stacking, i.e.,   monoclinic versus rhombohedral stacking. They correspond to the stacking configurations present in the bulk-CrI$_{3}$ high-temperature (monoclinic) and low-temperature  (rhombohedral) phases, respectively. Bulk CrI$_{3}$ has as stable phase the rhombohedral structure (with R$\overline{\rm 3}$ space group) below $\sim$200~K, and the monoclinic structure (C$_{2h}$ space group) at higher temperatures.\cite{McGuire2015} 

The bilayer with monoclinic stacking (C$_{2h}$ symmetry) is the common form obtained by exfoliation from bulk CrI$_{3}$ at room temperature, and remains experimentally in that structure when cooled to low temperatures.\cite{SunYiSon19,Ubrig_2019} The bilayer in that structure has been the focus of most attention due to its layered antiferromagnetic (AFM) ground state (below T$_N\approx 45$~K), which can be conveniently switched to the FM state (by electric or magnetic means).\cite{HuangNature2017,HuaClaKle18,Jiang_Nat_Nanotech_2018}
The bilayer with rhombohedral  stacking (S$_{6}$ symmetry) is obtained experimentally by layer-stacking engineering, and is FM.\cite{Song_2019,Li_2020} 
Although in this paper we mainly focus on the spin texture of the FM bilayer with monoclinic stacking, we will also show that virtually the same spin texture pattern is present in the FM bilayer with rhombohedral stacking.

The two experimental structures of the bilayer correspond to configurations in which the two CrI$_3$ monolayers have the same chirality,\cite{Gibertini_2020} as identified by the directions of the triangles corresponding to the top facets of the iodine octahedra (see solid triangles in Fig.~\ref{structure}(b)) ---or equivalently by the upward oriented directions of the body diagonals of the octahedra (see blue arrows in Fig.~\ref{structure}(a) and (b)). 
Other meta-stable closed-packed structures of the CrI$_3$ bilayer have been predicted theoretically, and include structures in which the two layers have opposite chirality.\cite{Gibertini_2020} The later structures were shown to be FM.\cite{Gibertini_2020}

Unlike the two experimental structures, the non-chiral bilayer structures\cite{Gibertini_2020} are characterized by a horizontal mirror-reflection plane, $M_{xy}$, which transforms the bilayer into itself or a translation of itself. In the following we define as `'chiral'' structures, the atomic geometries lacking such symmetry (which include the two experimental forms of the bilayer as well as the monolayer). We note that we use this definition strictly in the context of FM crystals with perpendicular magnetization considered in this work. In that case, for a given chosen spin magnetization orientation $\hat{z}$, e.g., upward in Fig.~\ref{structure}(a), the chiral bilayer or monolayer structure and its horizontal mirror reflection are non-superposable  (as left-handed and right-handed) configurations.

There are three particular  directions related to the structural chirality of the bilayer (and of the monolayer), which correspond to the oriented diagonals of the Iodine octahedra (with directions pointing from bottom to top I-layer atom) ---one of these directions is the $z'$ axis indicated in Fig.~\ref{structure}(a). These three directions have the same tilt angle ($\theta=55.4^\circ$) with respect to the vertical $z$ axis and are enclosed in three vertical planes rotated by $120^0$ from each other about the $z$ axis. The projections of these three directions on the horizontal $xy$ plane are indicated (blue arrows) in Fig.~\ref{structure}(b).

The CrI$_3$ single monolayer is FM below T$_C = 45$~K (with D$_{3d}$ atomic-structure symmetry). 
The relativistic band structures of the FM monolayer and bilayers have been reported in previous studies.\cite{AFM_SG,SG_2021,Lado_2017,Jiang_PRB_2019}  
The CrI$_3$ monolayer exhibits an upper valence band which is isolated in energy over almost the whole BZ ---except at the BZ-edge K point (see Appendix). This band has a spin polarization which is virtually parallel to the Cr spin moments,\cite{AFM_SG} except near the BZ edge,  and displays thus a negligible in-plane spin texture in the main central part of the BZ (see  Appendix). In the bilayer, on the contrary, as will be shown in this study, the upper valence band displays an interesting and intense in-plane spin texture. This makes the CrI$_3$ bilayer a convenient system to study the effects of interlayer interactions on the spin texture of a vdW FM bilayer. In the following, unless otherwise mentioned, the results shown are for the CrI$_3$ FM bilayer with monoclinic  stacking.

The magnetic point group of the FM bilayer with monoclinic stacking is  C$_{2h}$[C$_i$] and  corresponds to the symmetry operations: i) identity E, ii) inversion I, iii) 180$^o$ rotation about the y axis combined with time reversal R$_{y}$[180$^0$]$\cdot$T, and iv) $xz$ mirror-plane reflection combined with time reversal M$_{xz}\cdot$T. 
The magnetic point group of the single FM monolayer is D$_{3d}$[S$_6$], while  that of the FM bilayer with rhombohedral stacking is just S$_{6}$.

\section{Computational methods}
\label{sec:method}

Our calculations are performed using density functional theory (DFT) as implemented in the Quantum ESPRESSO package with a plane-wave basis set.\cite{GiaBarBon09} The exchange-correlation interactions between electrons are treated within the local spin density approximations (LSDA).\cite{Perdew_Zunger,PRL_1980} Spin-orbit coupling is included within the (non-collinear) spinor DFT formalism using fully-relativistic projector-augmented-wave (PAW) pseudopotentials.\cite{kucukbenli2014projector,PhysRevB.50.17953} The plane-wave kinetic energy cutoffs for the electronic orbitals and charge density are chosen to be 60 Ry and 650 Ry, respectively. The BZ is sampled using the uniform Monkhorst-Pack $k$-point grid of $24 \times 24 \times 1$. The periodic images of the CrI$_{3}$ bilayer are separated by a vacuum distance of $\sim$30 \AA\ along the out-of-plane direction (same for the monolayer). During structural relaxations atomic coordinates are relaxed until the forces on atoms become less than $1 \times 10^{-4}$ Ry/Bohr.

The calculated in-plane lattice constant of the CrI$_{3}$ bilayer is $a = 6.69$~\AA, while the equilibrium distance $d$ ($d_{\rm int}$) between the two outermost (innermost) iodine atomic layers of the bilayer is $d= 9.47$~\AA\ ($d_{\rm int}^{\rm eq}=~3.36$~\AA), consistent with previous LDA calculations.\cite{Leon_2020,SG_2021}  Within each  CrI$_{3}$ monolayer, the vertical distance between the Cr plane and the innermost I layer ($d_{1}$) and outermost I layer ($d_{2}$) differ slightly (by $\sim 0.1\%$) in the bilayer configuration ($d_{1} < d_{2}$), breaking the local inversion symmetry within each monolayer, while for the monolayer $d_{1}=d_{2}$. 

For the strained bilayer, we apply a vertical compressive strain $\epsilon_{zz} = -5$\%, as described in Ref.\citenum{SG_2021}. The optimized geometry is obtained by fixing the $z$ coordinates of the outermost I atomic planes to the distance $d$ contracted by 5\%, while all other atomic coordinates are free to relax, and the in-plane lattice parameter is also optimized. The 5\%  compression expands the in-plane lattice parameter by $1\%$ and reduces $d_\mathrm{int}$ by $ 9\%$. The magnetic anisotropy energies of the pristine and strained bilayer are found to be 0.65~meV/Cr and 0.36~meV/Cr, respectively,\cite{SG_2021} establishing the fact that out-of-plane orientation of the Cr spin is favored in both cases.

We determine the spin texture of the bands in k-space by calculating the expectation values of the spin operators  ${\frac{1}{2}}\hat{\sigma}_{\alpha}$, $\alpha = x,y,z$, in the Bloch spinor eigenfunctions $\Psi_{n, {\bm k}}({\bm r})$:  
\begin{equation}
S_{\alpha}(n,{\bm k}) = \frac{1}{2}  {\frac{\langle\Psi_{n, {\bm k}}|\hat{\sigma}_{\alpha}|\Psi_{n, {\bm k}}\rangle }{\langle\Psi_{n, {\bm k}}|\Psi_{n, {\bm k}}\rangle}},
\end{equation}
where $n$ is the band index and $\hat{\sigma_{x}}$, $\hat{\sigma_{y}}$,  and $\hat{\sigma_{z}}$ are the Pauli matrices. To present the in-plane spin texture of the band, we display the spin-polarization vector ${\bm S_{\parallel}}({\bm k}) = (S_{x}({\bm k}), S_{y}({\bm k}))$ at $k$-grid points of the 2D Brillouin zone (BZ), and for the out-of-plane spin texture, we show the map of $S_{z}$ isovalues in the 2D BZ.

For some specific $\Psi_{n, {\bm k}}$ states, we also examine the corresponding $\Psi_{n, {\bm k}}$--state spin-polarization density, ${\bm m}^{(n, {\bm k})}({\bm r})$, whose components are calculated as:  
\begin{equation}
m_{\alpha}^{(n,{\bm k})}({\bm r}) = \mu_B \Psi^{+}_{n, {\bm k}}({\bm r}) \hat{\sigma}_{\alpha} \Psi_{n, {\bm k}}({\bm r}), 
\end{equation}
where $\mu_B$ is the Bohr magneton.

We note that for centro-symmetric ferromagnets, although time reversal symmetry is broken, space-inversion symmetry still commutes with the Kohn-Sham Hamiltonian, H$_{KS}$. For the eigenvalues this implies: E$_n(-{\bm k})$ = E$_n({\bm k})$, while for the non-degenerate eigenstates at ${\bm k}$ (the typical case in a ferromagnet) this implies:   ${\bm m}^{(n,-{\bm k})}({\bm r}) = {\bm m}^{(n, {\bm k})}(-{\bm r})$ and $ {\bm S}(n, -{\bm k}) = {\bm S}(n, {\bm k})$. 



\section{Spin texture in pristine FM bilayer} 
\label{sec:pristine}

\begin{figure*} [t]
\includegraphics[width=0.85\textwidth] {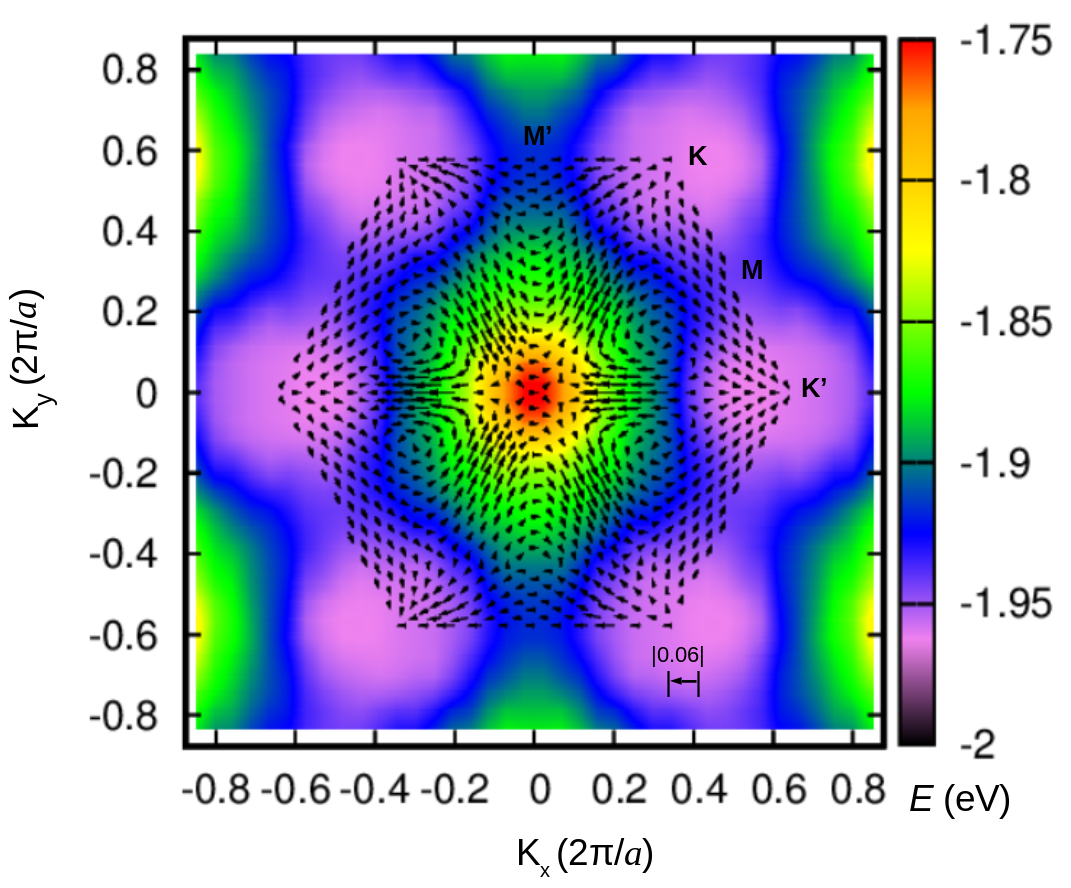}
\caption{In-plane spin texture plot in the 2D BZ for the highest valence band of the pristine FM CrI$_{3}$ bilayer. The magnitude of the in-plane spin component,  ${\bm S}_{\parallel}({\bm k})$, is proportional to the length of the arrow, according to the scale indicated in the bottom-right corner of the plot.  The color map indicates the energies of the upper valence band in $k$-space.}
\label{FM_firstVS_spint}
\end{figure*}

In Fig.~\ref{FM_firstVS_spint}, we display the in-plane spin texture for the highest valence band of the FM bilayer (with monoclinic stacking) in the 2D BZ. The energy-isovalue map of the band in $k$-space is also shown in this figure, for easy reference ---with the valence band maximum (VBM) at $\Gamma$ clearly visible.  Contrary to the FM monolayer, whose in-plane spin texture is minute for the upper valence band (Fig.~\ref{ml_st} in  Appendix), for the bilayer the highest valence band displays significant spin canting away from the $z$ axis, giving rise to a pronounced and interesting in-plane spin texture in the central part of the  BZ. 

The in-plane texture, in Fig.~\ref{FM_firstVS_spint}, has as dominant features six intense ``spin patches'', each made of strong parallel spins. These patches are located around $\Gamma$, along the $\Gamma-K'$, $\Gamma - K$, and equivalent directions, and have alternating inward and outward radial-spin directions. The patches do not correspond to any particular points in the band's dispersion. One can notice, in between patches, around $\Gamma$, six alternate vortex structures with centers located at slightly larger distance from the BZ  center, along the $\Gamma-M'$, $\Gamma - M$, and equivalent directions. All of these features are absent for the highest valence band of the monolayer, which exhibits a noticeable (but small) in-plane texture essentially only along the edge of the BZ, near the M point (Fig.~\ref{ml_st}). Although a similar BZ-edge spin pattern is also present for the highest valence band of the bilayer (with a slightly decreased intensity), it is barely visible on the scale of the bilayer spin texture in Fig.~\ref{FM_firstVS_spint}.

Proceeding thus from the monolayer to the bilayer causes intense in-plane spin patch features to appear in the spin texture of the highest valence band around $\Gamma$. We note that these spin-texture features are not connected to any structural relaxation of the monolayer due to the bilayer formation from isolated monolayers. This has been verified by constructing a bilayer with the same frozen monolayer Cr-I inter-planar distances as in the isolated monolayer, which gives a spin texture (see SI) virtually identical to that in Fig.~\ref{FM_firstVS_spint}. 
Hence, this indicates that the spin patches, in Fig.~\ref{FM_firstVS_spint}, are a direct effect of  the interaction between the two monolayers on the electronic spin texture, which is a rather intriguing effect.

    In addition, we find that when the interlayer stacking is changed from the monoclinic stacking (C$_{2h}$) to the rhombohedral stacking (S$_{6}$), which corresponds  to the other  experimentally known phase of the bilayer, almost the same in-plane spin texture is obtained for the upper valence band (see SI). Thus, the different relative-lateral translation of the CrI$_3$ monolayers, which differentiate the two structural phases, has only a minor influence on the in-plane spin texture. In fact, for the rhombohedral stacking, both the intensity of the in-plane spin texture  $|{\bm S_{\parallel}}({\bm k})|$ and the band energy E$({\bm k})$ exhibit perfect hexagonal symmetry in the BZ, while for the monoclinic stacking, one can observe in the central part of the BZ, in Fig.~\ref{FM_firstVS_spint}, a slight uniaxial elongation along the $k_y$ relative to $k_x$ axis of both the spin-texture pattern and band-energy map. Such a smooth uniaxial deformation is a direct consequence of the monoclinic stacking characterized by a 2D macroscopic uniaxial anisotropy of the crystal structure along the $y$ relative to $x$ axis in the bilayer  [see Fig.~\ref{structure}(b)].

\begin{figure*}[t]
		\includegraphics[width=0.7\textwidth]{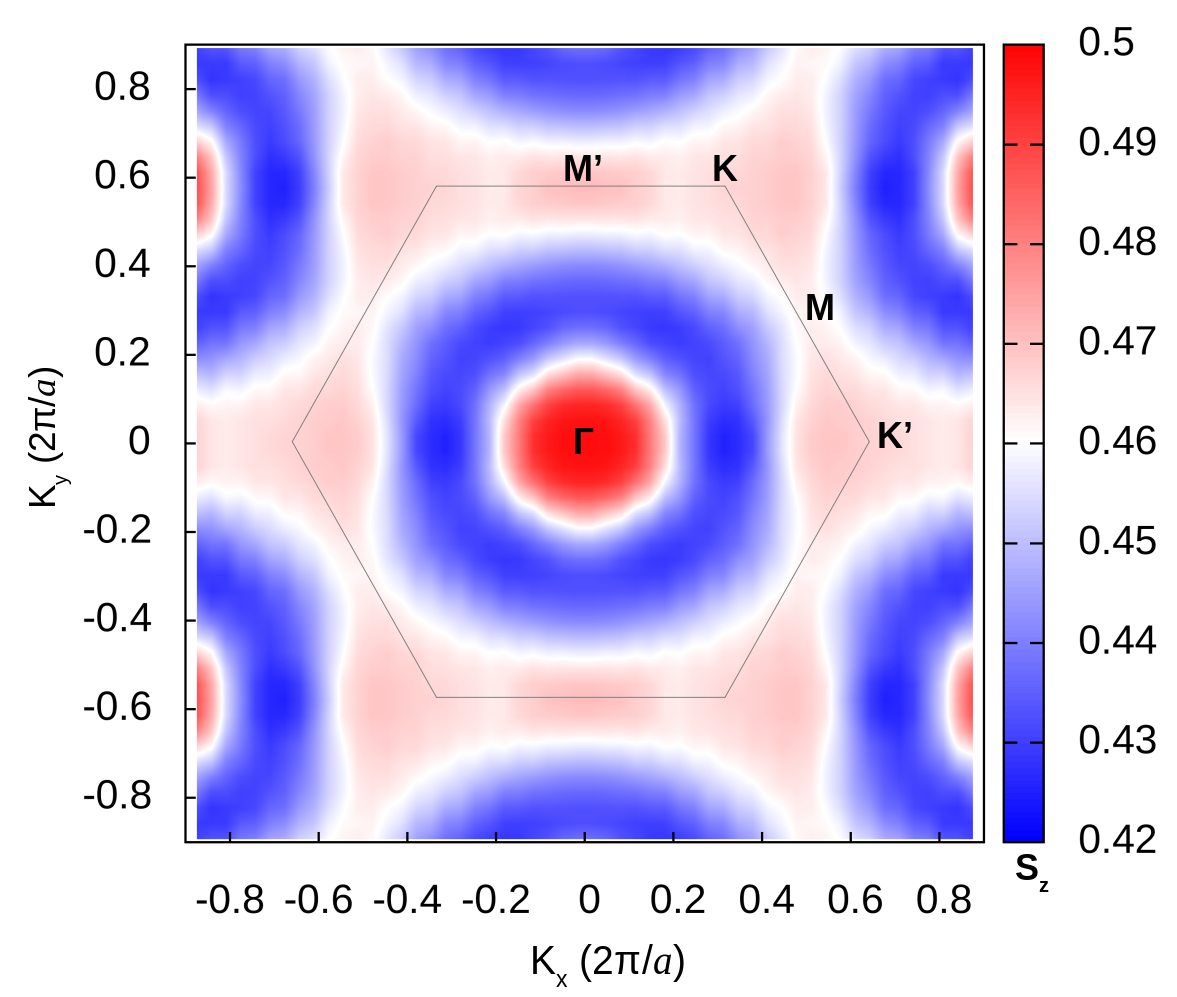}	
        \caption{\label{Pristine_Sz} Isovalue map of the out-of-plane spin component $S_z$ in k-space for the upper valence band of the CrI$_{3}$ FM bilayer. The value of $S_z$ is represented by the color scheme indicated on the right of the figure.} 
\end{figure*}

For the states of the highest valence band, the maxima in the amplitude of ${\bm S_{\parallel}}({\bm k})$ correspond to the six spin patches. The largest amplitude is obtained for the patches along the $k_x$ axis, 
at $ k_x = \pm 0.23 \cdot 2\pi/a$, 
and is 0.065. The amplitude of ${\bm S_{\parallel}}({\bm k})$ is vanishing at the $\Gamma$ point, at the centers of the vortices, and at the K and K' corners of the BZ. 
In Fig.~\ref{Pristine_Sz}, we show the values of the out-of-plane spin component $S_{z}$ of the upper valence band in the 2D BZ. Similar to the layer-resolved spin texture of the CrI$_3$ AFM bilayer,\cite{AFM_SG} a ring-type texture is observed in the $S_z$ of the FM bilayer.  In Fig.~\ref{Pristine_Sz} the largest value is at the BZ centre and is 0.50; $S_z$ decreases then with increasing $k$ radius, reaching a minimum value of 0.43 at $k_x = \pm 0.3 \cdot 2\pi/a $. For larger radii, $S_z$ smoothly increases with increasing radius up to the BZ edge, where it exhibits local maxima at $K$, $K'$, $M$, and $M'$. At the points where $|{\bm S_{\parallel}}|$ is the largest, its value is 15\% of the amplitude of the out-of-plane spin component.  We note that the vanishing in-plane spin components at $\Gamma$  and at $K$,  $K'$ points correspond to the maximum and local maxima of 
$S_z$, respectively.

Similar to the highest valence band, we find that the second highest valence band also has the strongest ${\bm S_{\parallel}}({\bm k})$ components located in spin-patch regions centered along the $\Gamma - K$ and $\Gamma - K'$ directions (see Fig.~\ref{VS_1_bl} in Appendix). For the second-highest valence band, however, the ${\bm S_{\parallel}}({\bm k})$ components in the patches are aligned along the opposite direction in $k$-space with respect to those of the highest valence band, and there is no vortex. The spin patterns along the BZ edge (and in particular near $M$ and $M'$) are similar, instead, for the two bands, and have the same spin orientation, while the absence of vortices allows much larger values of the in-plane spins, close to the $M'$ points. This BZ-edge pattern is a feature  originating from the spin texture of the monolayer highest-valence band (see Fig.~\ref{ml_st}), which persists upon formation of the bilayer in the spin textures of its upper two valence bands.

For the ${\bm S_{\parallel}}({\bm k})$ of the highest valence band, in Fig.~\ref{FM_firstVS_spint}, we note that the direction of the pre-existing (monolayer) BZ-edge spins along the $K - K'$ line and the directions of the spins in the patches along the $K' - \Gamma$ and $\Gamma - K$ lines are oriented clockwise around the $K' - K - \Gamma - K'$ triangular contour. This generates (by continuity) a vortex centered on the diagonal, $\Gamma$ - $M$, of the triangle. For the bilayer's second-highest valence band, instead, the spins of the patches, along  $K' - \Gamma$ and along $\Gamma - K$, are in the reversed directions, and their resultant has the same orientation as the pre-existing spins along the $K - K'$ BZ edge (Fig.~\ref{VS_1_bl}). This produces an extended region of parallel spins (around $M$, and similarly around $M'$) and no vortex. Hence, the presence of the vortex features, in Fig.~\ref{FM_firstVS_spint}, (and their absence in the spin texture of the second valence state in Fig.~\ref{VS_1_bl}) is a consequence of the appearance of the patch features and their specific spin orientations. 

Interestingly, compared to the spin textures of nonmagnetic crystals, where the rules are that for ${\bm k}$ parallel to a mirror plane M$_i$ of the space group, one has ${\bm S}({\bm k}) \perp$ M$_i$, and for  ${\bm k}$ parallel to a rotation axis R$_i$, one has ${\bm S}({\bm k}) \parallel$ R$_i$,\cite{Zunger_2021} here we have just the reversed behavior.  In Fig.~\ref{FM_firstVS_spint},  we see that for ${\bm k}$ parallel to the R$_y$ axis, we have ${\bm S}({\bm k}) \perp$ R$_y$, and 
for ${\bm k}$ parallel to the M$_{xz}$ plane, we have ${\bm S}({\bm k}) \parallel$ M$_{xz}$. 
As will be shown in Section~\ref{sec:sym}, this corresponds to two general rules for operators of a centrosymmetric system's magnetic group that are combination of time reversal with a rotation operation, R$_i\cdot$T, or with a mirror-reflection operation, M$_i \cdot$T, namely: i) for ${\bm k} \parallel R_i$, one has ${\bm S}({\bm k}) \perp$ R$_i$, ii) 
for ${\bm k} \parallel M_i$, one has  ${\bm S}({\bm k}) \parallel$ M$_i$. In centrosymmetric systems, the rules are in fact reversed for rotation and reflection operations combined with time reversal, relative to the rules established for operators not involving time reversal.

\section{Planar-averaged spin-polarization densities in pristine FM bilayer} 
\label{sec:profile}

\begin{figure*}[t]
		\includegraphics[width=0.7\textwidth]{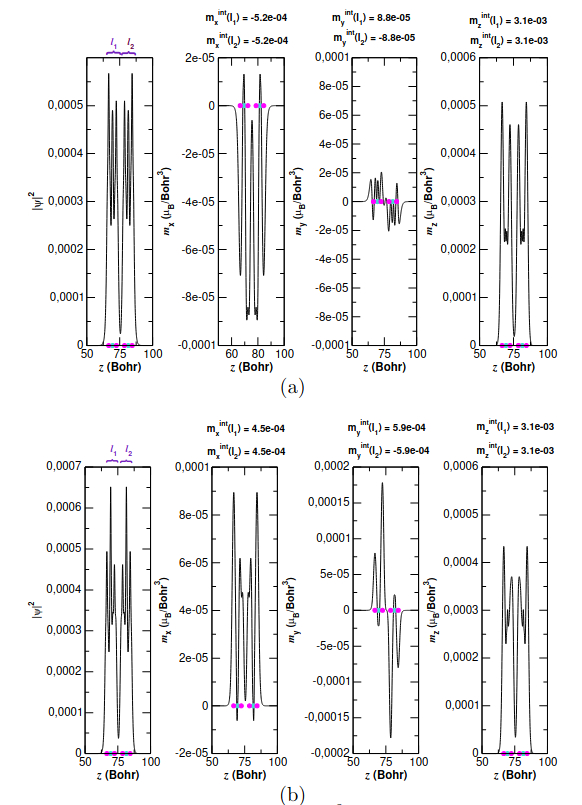}	
        \caption{\label{magn_directn1}Planar average of the probability density $|\Psi|^{2}$ and spin-polarization density components $m_{x}$, $m_{y}$ and $m_{z}$ calculated at ${\bm k}= (0.2, 0) \cdot 2 \pi/a$ for the highest valence state (a) and second-highest valence state (b) of the FM bilayer. The turquoise and magenta circles show the positions of Cr and I atoms, respectively, along the $z$ direction of the supercell;  $m_{\alpha}^\mathrm{int}(\mathrm{l_{1}})$ [$m_{\alpha}^\mathrm{int}(\mathrm{l_{2}})$], $\alpha = x,y,z$, are obtained by integrating $m_{\alpha}(z)$ over half of the supercell length in the z-region starting [ending] at the mid-point between the two layers, l$_{1}$ and l$_{2}$, and have units of $\mu_{B}/\mathrm{Bohr^{2}}$. } 
\end{figure*}

\begin{figure*}[t]
		\includegraphics[width=0.7\textwidth]{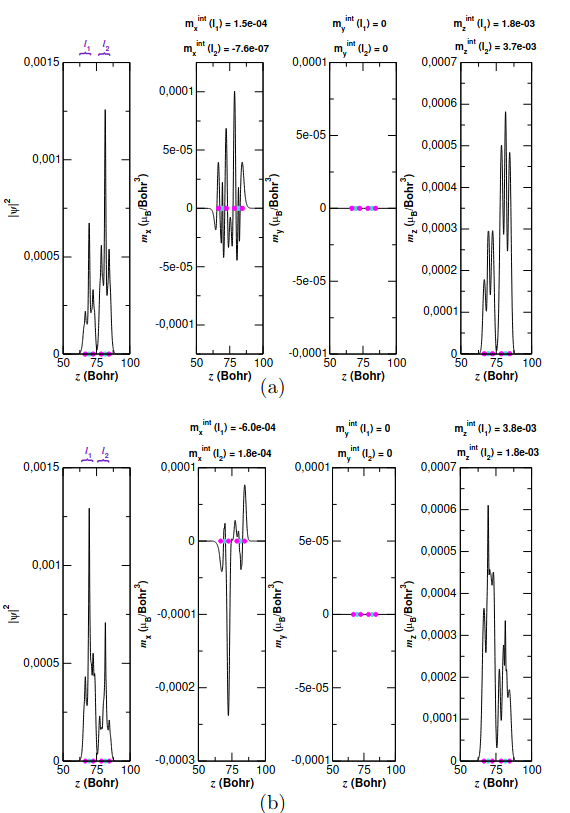}	
        \caption{\label{magn_directn2}Same as in Fig.~\ref{magn_directn1}, but calculated at ${\bm k}= (0, 0.3) \cdot 2 \pi/a$. } 
\end{figure*}

In Figs.~\ref{magn_directn1} and \ref{magn_directn2}, we display the planar averages of the probability density, $|\psi|^2(z)$, and spin-polarization density components, $m_x(z)$, $m_y(z)$, $m_z(z)$, for the highest and second-highest valence state at two different ${\bm k}$ points chosen at similar distances from the BZ center, one on the $k_x$ axis [Fig.~\ref{magn_directn1}] and the other one on the $k_y$ axis [Fig.~\ref{magn_directn2}]. On the $k_x$ axis, we took the point at $ k_x = 0.2 \cdot 2\pi/a$, which is located within a ${\bm S_{\parallel}}({\bm k})$-patch region (see Fig.~\ref{FM_firstVS_spint}). On the $k_y$ axis, we chose the point at $k_y = 0.3 \cdot 2\pi/a$,  positioned in between two ${\bm S_{\parallel}}({\bm k})$-patch regions (see Fig.~\ref{FM_firstVS_spint}), out of which one with radial outward spin at  ${\bm k_+} = (0.2, 0.3) \cdot 2\pi/a$ and one with radial inward spin at ${\bm k_-} = (-0.2, 0.3)\cdot 2\pi/a$.  The chosen point on the $k_y$ axis  is observed to have the resultant (small) spin ${\bm S_{\parallel}}$  oriented   in the $x$--direction. 

For ${\bm k}$ along the $k_x$ axis, in Fig.~\ref{magn_directn1}, the profiles of the  probability density, $|\psi|^2(z)$, and of the spin-polarization-density components  $m_x(z)$ and $m_z(z)$ are symmetric within the bilayer, i.e., with respect to the $z$-inversion at the midpoint between the two layers, $l_1$, $l_2$, of the bilayer.  The $m_x(z)$ radial components (along ${\bm k}$) of the two states are seen to have similar amplitude and opposite signs, consistent with the sign reversal found for the patch  spins ${\bm S_{\parallel}}({\bm k})$ of the two bands. In addition, in Fig.~\ref{magn_directn1}, we observe for both states  the presence of in-plane tangential $m_y(z)$ components of identical magnitude and opposite sign on $l_1$ and $l_2$, which do not contribute to the ${\bm S_{\parallel}}({\bm k})$ of the states. The magnitude of $m_y(z)$ is largest for the second state, and is comparable in that case to the magnitude of $m_x(x)$. 

The in-plane tangential $m_{y(z)}$ components reverting sign from $l_{1}$ to $l_{2}$,  in Fig.~\ref{magn_directn1}, are consistent with a Rashba effect induced by the local vertical electric field, $\mathbfcal{E}_z^{(i)}$, on each layer $l_{i}$, due to the presence of the other layer. The fields $\mathbfcal{E}_z^{(i)}$ are opposite on the two layers, and may be viewed in the phenomenological Rashba model as local magnetic effective fields  ${\bm B}_{\rm eff}^{(i)} \sim {\bm k} \times  \mathbfcal{E}_z^{(i)}$ canting the spin polarization vectors in opposite directions on the two layers. In the AFM CrI$_3$ bilayer, similar in-plane tangential components of the spin polarization were found for the two highest (degenerate) valence states for ${\bm k}$ around the BZ center.\cite{AFM_SG} In the latter case, however, the two states were segregated, each on one of the layer, and displayed opposite spin-polarization components. Here, instead, for the FM bilayer, the two opposite in-plane spin-polarization components on the two layers are found to co-exist in the same non-degenerate state. 


For ${\bm k} = (0,0.3) \cdot 2\pi/a$, in Fig.\ref{magn_directn2}, the planar averages reveal a striking asymmetry of the probability density $|\psi|^2(z)$ and of the out-of-plane spin-polarization component $m_z(z)$ on the two layers. They both exhibit a substantially larger weight on layer $l_2$ than on layer $l_1$ for the highest valence state, and the reversed trend for the second-highest valence state. Such an asymmetry is allowed by symmetry in our system at ${\bm k}$ points with $k_y \neq 0$ (as will be seen in Section~\ref{sec:sym}), but is unlike the behavior in nonmagnetic centrosymmetric crystals and in FM  centrosymmetric systems without SOC. For ${\bm k} \parallel k_y$, in Fig.\ref{magn_directn2}, we also note that the $m_y(z)$ component is zero, and $m_x(z)$ is the only in-plane spin-polarization component present. 

The $m_x(z)$ component in Fig.~\ref{magn_directn2} is neither symmetric nor antisymmetric, and appears to be the outcome of an interference for both states.  This, in fact, can explain what is the mechanism controlling the weight difference in the  probability density, i.e., an interference between the Rashba and non-Rashba effect. Indeed, we have seen that for ${\bm k} = (0,0.3) \cdot 2\pi/a$, the ${\bm S_{\parallel}}$ for the upper valence band is along the $k_x$ axis and has positive small $S_x$ component (see Fig.~\ref{FM_firstVS_spint}), resulting from the continuation of spins between  the two neighbouring non-Rashba  spin patches   ${\bm S_{\parallel}}({\bm k_{\pm}})$. The corresponding non-Rashba spin-polarization-density component, denoted $m^{\rm NR}_x(z)$, has therefore to be globally (over the whole bilayer) positive for the upper valence state. Similarly, for the second state, with opposite orientation of ${\bm S_{\parallel}}$ (see  Fig.~\ref{VS_1_bl}), the $m^{\rm NR}_x(z)$ component must be globally negative. At the same time, for ${\bm k}$ along the $k_y$ axis the Rashba-induced spin-polarization-density component is also expected to be in the $x$--direction. However, the corresponding $m^{\rm R}_x(z)$ should be opposite on the two layers, i.e, negative on $l_1$ and positive on $l_2$ for both states (considering the orientation of the local electric fields  $\mathbfcal{E}^{(1)}$ and $\mathbfcal{E}^{(2)}$, inferred from the Rashba $m_y(z)$ components at the other ${\bm k}$ point, in Fig.~\ref{magn_directn1}). 
Hence for the first state, the Rashba $m^{\rm R}_x(z)$ and the non-Rashba $m^{\rm NR}_x(z)$ components are in phase on layer $l_2$ (both globally positive) and in counter-phase on layer $l_1$ (cancelling contributions), whereas for the second state, they are in phase on $l_1$ (globally negative) and in counter-phase on $l_2$ (mostly cancelling contributions), consistent with the observed profiles of the final $m_x(z)$ components in Fig.\ref{magn_directn2}.

Furthermore, in the case of the first state, in Fig.\ref{magn_directn2}, the constructive interference of the Rashba effect and non-Rashba effect in $m_x(z)$ on layer $l_2$ and destructive one on layer $l_1$ are expected to favor (energetically) residence of that state on $l_2$, shifting the probability-density weight to that layer. For the second state, instead, the constructive interference is on $l_1$, shifting the weight to that layer. This is in agreement with the trends in the weights of $|\psi|^2(z)$ and $m_z(z)$ observed, in Fig.~\ref{magn_directn2}, for both states.


\section{Implications of ferromagnet symmetries on spin texture}
\label{sec:sym}

In order to better understand the pattern of the spin texture as well as the spatial behavior of the probability density and spin-polarization densities in our prototype FM system, we have addressed their symmetry properties.  We focus on non-degenerate states at ${\bm k}$ points around $\Gamma$, i.e., with ${\bm k}$ relative to the BZ center. To establish in general symmetry properties considering the operators of the magnetic group of a ferromagnet, we distinguish between operators involving time reversal and those which do not involve time reversal. The correspondence relations for the  quantities ${\bm S}$, ${\bm m}$, and $|\psi|^2$ at ${\bm k}$ points related by these different types of symmetry operators of a magnetic group are given in the Appendix. 

In general, for an operator $\{R|{\bm f}\}$ that does not involve time reversal,  i.e., where $R$ is a spatial point-group operation and ${\bm f}$ a fractional translation,\cite{Bassani_1975} the symmetry correspondence relations are between ${\bm k}$ and $R{\bm k}$, whereas for an operator  $\{R'|{\bm f'}\}T$ combining a spatial operation $\{R'|{\bm f'}\}$ and time reversal, the symmetry relations are between ${\bm k}$ and $-R'{\bm k}$ (see appendix \ref{app:symmetries}). In particular, for the spin expectation values ${\bm S}$ the relations are: I)  ${\bm S}(R{\bm k}) = R{\bm S}({\bm k})$ for an operator that does non involve  time reversal, and II) 
  ${\bm S}(-R'{\bm k}) = R'T{\bm S}({\bm k})$ for an operator involving time reversal.

The spin expectation values ${\bm S}$ and the spin-polarization density  ${\bm m}$ are axial vectors (vectors with a virtual current loop) and transform accordingly under time reversal and under spatial operations $R$, $R'$.\cite{Chapon_2012} Taking this into account and using the operators of the magnetic group of the FM bilayer (see Section \ref{sec:system}) in the relations reported in the Appendix yields  the correspondence rules, given in 
Table~\ref{Symm_oper}, for the probability density, spin-polarization density, and spin expectation values in the FM bilayer at ${\bm k}$ points related by the symmetry operators.

\begingroup
\begin{table*}[h]
{
\caption{Symmetry operators of the magnetic point group of the CrI$_{3}$ FM bilayer and correspondence rules for the probability density, spin-polarization density, and spin expectation values at ${\bm k}$ points related by the symmetry operators. The reference probability-density function  $|\psi|^2({\bm r})$, spin-polarization-density functions   $m_x({\bm r})$,  $m_y({\bm r})$,  $m_z({\bm r})$, and spin expectation values $S_x$, $S_y$, $S_z$ are those of a non-degenerate eigenstate at a point $(k_x,k_y)$ in the BZ. The position ${\bm r}$ is relative to an inversion symmetry point of the bilayer, and ${\bm k}$ is relative to the BZ center.}

\begin{tabular}{|c|c||c|c|c|}
\hline
Symmetry & ${\bm k}$ & $|\Psi_{{\bm k}}|^{2}$ & ${\bm m}_{{\bm k}}$ &${\bm S}_{{\bm k}}$  \\ 
\hline 
E & $k_{x}, k_{y}$ & $|\Psi|^{2}$(x, y, z) & $m_{x}$(x, y, z), $m_{y}$(x, y, z), $m_{z}$(x, y, z)& $S_{x}, S_{y}, S_{z}$ \\
\hline
I & $-k_{x}, -k_{y}$ & $|\Psi|^{2}$(-x, -y, -z) & $m_{x}$(-x, -y, -z), $m_{y}$(-x, -y, -z), $m_{z}$(-x, -y, -z)& $S_{x}, S_{y}, S_{z}$ \\ 
\hline
R$_{y}$(180$^0$)$\cdot$T & $k_{x}, -k_{y}$ & $|\Psi|^{2}$(-x, y, -z) & $m_{x}$(-x, y, -z), $-m_{y}$(-x, y, -z), $m_{z}$(-x, y, -z)& $S_{x}, -S_{y}, S_{z}$ \\ 
\hline
M$_{xz}\cdot$T & $-k_{x}, k_{y}$ & $|\Psi|^{2}$(x, -y, z) & $m_{x}$(x, -y, z), $-m_{y}$(x, -y, z), $m_{z}$(x, -y, z)& $S_{x}, -S_{y}, S_{z}$ \\ 
\hline

\end{tabular}
\label{Symm_oper}
}
\end{table*}
\endgroup

The symmetry operations $E$ and $I$, in Table~\ref{Symm_oper}, yield for the spin polarization vector, ${\bm S}$, as expected:  ${\bm S}(-{\bm k}) = {\bm S}({\bm k})$. 
Furthermore, considering the symmetry operators $I$ and $M_{xz} T$, in Table~\ref{Symm_oper}, the correspondence rule on ${\bm S}$ implies that for  ${\bm k}$ points belonging to the $M_{xz}$ plane ($k_y = 0$), the $S_y$ component must be zero, and therefore ${\bm S}({\bm k}) \parallel$ M$_{xz}$.  Similarly, considering the operators $I$ and R$_{y}($180$^0)T$, in Table~\ref{Symm_oper}, the corresponding rule on ${\bm S}$ imposes that for  ${\bm k}$ points on the R$_y$ axis $S_y$ must be zero, and thus ${\bm S}({\bm k}) \perp$ R$_y$. This is in agreement with the behavior noted in Section~\ref{sec:pristine}, where the two rules were  anticipated.  

Conspicuously, for non-degenerate states with wavevector  ${\bm k}$, the above relation (II) ${\bm S}(-R'{\bm k}) = R'T{\bm S}({\bm k})$ is valid in general for any operator $\{R'|{\bm f'}\} T$ of a system's magnetic group. More generally therefore, with 3D ${\bm k}$ vectors and any operators $\{M_i|{\bm f'}\} T$ and $\{R_i|{\bm f''}\} T$ of the magnetic group of a centrosymmetric system that are, respectively, reflection and rotation operations combined with time reversal, the relation (II) yields that for ${\bm k} \parallel M_i$,  ${\bm S}({\bm k}) \parallel$ M$_i$, and for  ${\bm k} \parallel R_i$,  ${\bm S}({\bm k}) \perp$ R$_i$. This is the reverse of the rules established for non-magnetic systems, where for ${\bm k} \parallel M_i$, one has ${\bm S}({\bm k}) \perp M_i$, and for ${\bm k} \parallel R_i$,  ${\bm S}({\bm k}) \parallel  R_i$, as obtained with operators not involving $T$. In centrosymmetric systems thus in general, time reversal is reverting those rules.

Furthermore, a notable consequence of the relation (I) is that for any CrI$_3$ bilayer structure which is non chiral (but has the same vertical FM alignment of the Cr spins),  the in-plane spin texture of the FM bilayer $S_{\parallel}({\bm k})$ must be zero at all ${\bm k}$ points. Indeed, such non-chiral FM bilayers are characterized by 
a horizontal mirror symmetry operator $\{M_{xy}|{\bm f}\}$ in their magnetic group. The corresponding relation (I) yields  $M_{xy}{\bm S}_{\parallel}({\bm k}) =$  ${\bm S}_{\parallel}(M_{xy}{\bm k}) = $ ${\bm S}_{\parallel}({\bm k})$ for ${\bm k}$ in the 2D BZ, and considering that an axial vector ${\bm S}_{\parallel}({\bm k})$ parallel to the mirror plane transforms as $M_{xy} {\bm S}_{\parallel}({\bm k}) = -{\bm S}_{\parallel}({\bm k})$, this implies that ${\bm S}_{\parallel}({\bm k}) = 0$.

Concerning the probability density, $|\Psi_{{\bm k}}|^{2}$,  the symmetry relation obtained with $E$ and $R_y(180^0)T$,  in Table~\ref{Symm_oper}, imposes that for ${\bm k}$ along the $k_x$ axis, the planar average, $|\psi|^2(z)$, must be symmetric under inversion through the $z$ mid-point between the two layers. However,  when $k_y \neq 0$, in Table~\ref{Symm_oper}, the probability-density weight may differ on the two layers, as we noticed in Section~\ref{sec:pristine}  (see also the layer-projected band structure in SI). This is in contrast to the case of non-magnetic centro-symmetric crystals, where the probability density at any ${\bm k}$ point $\rho_{{\bm k}}({\bm r})$   [given by  $\rho_{{\bm k}}(\overrightarrow{r}) = \sum_i|\psi_{i,{\bm k}}({\bm r})|^2$, with the sum over the set of degenerate states at ${\bm k}$] is symmetric under inversion    $\rho_{{\bm k}}({\bm r}) = \rho_{{\bm k}}(-{\bm r})$, due to the additional time-reversal invariance of the system.

\section{Mechanisms of interlayer-induced spin texture}
\label{sec:mechanism}

The symmetry rules of the previous section have explained some 
properties of the spin texture of the FM bilayer and have shown that the in-plane spin texture is related to the chirality of the two layers (as a whole, as well as the intrinsic chirality of each monolayer). 
These rules do not explain however the intensity behavior of the in-plane spin texture and the mechanism generating the spin-patch features with opposite spins for the two highest valence bands. In this section, we discuss the mechanisms of in-plane canting of the spin-polarization density ${\bm m}_{{\bm k}}$ on the two CrI$_3$ layers, caused by the interlayer interaction, and address the origin of the opposite patch spins for the two highest valence bands.

We have seen, in Sections~\ref{sec:pristine} and \ref{sec:profile}, that two effects of the interlayer interaction are present, around the BZ center, which cant the spin polarization  ${\bm m}_{{\bm k}}^{l_i}$ away from the $z$ axis on the layers $l_1$ and  $l_2$: the Rashba-related effect and the non-Rashba (chiral) effect, the latter leading  to the spin patches. 
For ${\bm k}$ along the $k_x$ axis, the Rashba effect induces opposite spin-polarization components $m_y^{l_1}$, $m_y^{l_2}$ on the two layers, for both states; whereas the other effect induces identical components $m_x^{l_1}$, $m_x^{l_2}$ on the two layers, but having opposite sign for the two states.  In the following, considering  different  components of the interaction between the layers, and based on the nature of the monolayer main states involved and their perturbation, we rationalize the behavior of the spin texture around  $\Gamma$, singling out the coupling between states that can account for the spin-patch behavior and for the Rashba behavior of the two upper valence states on the two layers. 

To do that we have first determined the main changes induced by the interlayer interaction in the highest valence states of the bilayer for ${\bm k}$ along the $\Gamma - K'$ and $\Gamma - K$ directions  (relevant for the spin patches). The relativistic band structure of the FM bilayer is plotted in Fig.~\ref{Jz_0.5_unstrained} and that of the FM monolayer in the Appendix (Fig.~\ref{ml_Jz_0.5}). The two highest valence bands of the bilayer have  dispersions resembling that of the monolayer. These bands can be viewed   as a superposition of the highest-valence band of the two isolated monolayers, with some additional $k$-dependent splitting and smooth rise in band energy away from the BZ center, caused by the interlayer interaction. 
For the monolayer, the valence-band edge in the central part of the BZ  corresponds to I-$5p$ spin-orbital $|J=3/2,J_z=+3/2\rangle$ states, with some small additional I-$5p$ $|J=3/2,J_z=-1/2\rangle$ component incorporated when moving away somewhat from $\Gamma$, i.e., for $k$ radii $0.2-0.3 \cdot 2 \pi/a$. The resulting states of the monolayer have vanishing in-plane spin  ${\bm S}_{\parallel}({\bm k})$, given the rule for non-zero matrix elements of the $\hat{\sigma_{x}}$ and $\hat{\sigma_{y}}$  operators: $\langle J, J_{z}|\hat{\sigma}_{\alpha}|J, J'_{z}\rangle \neq 0$ for $\alpha = x,y$, only when $\Delta J_z =J_z-J'_z=\pm 1$. 

\begin{figure*} [t]
\includegraphics[width=0.95\textwidth] {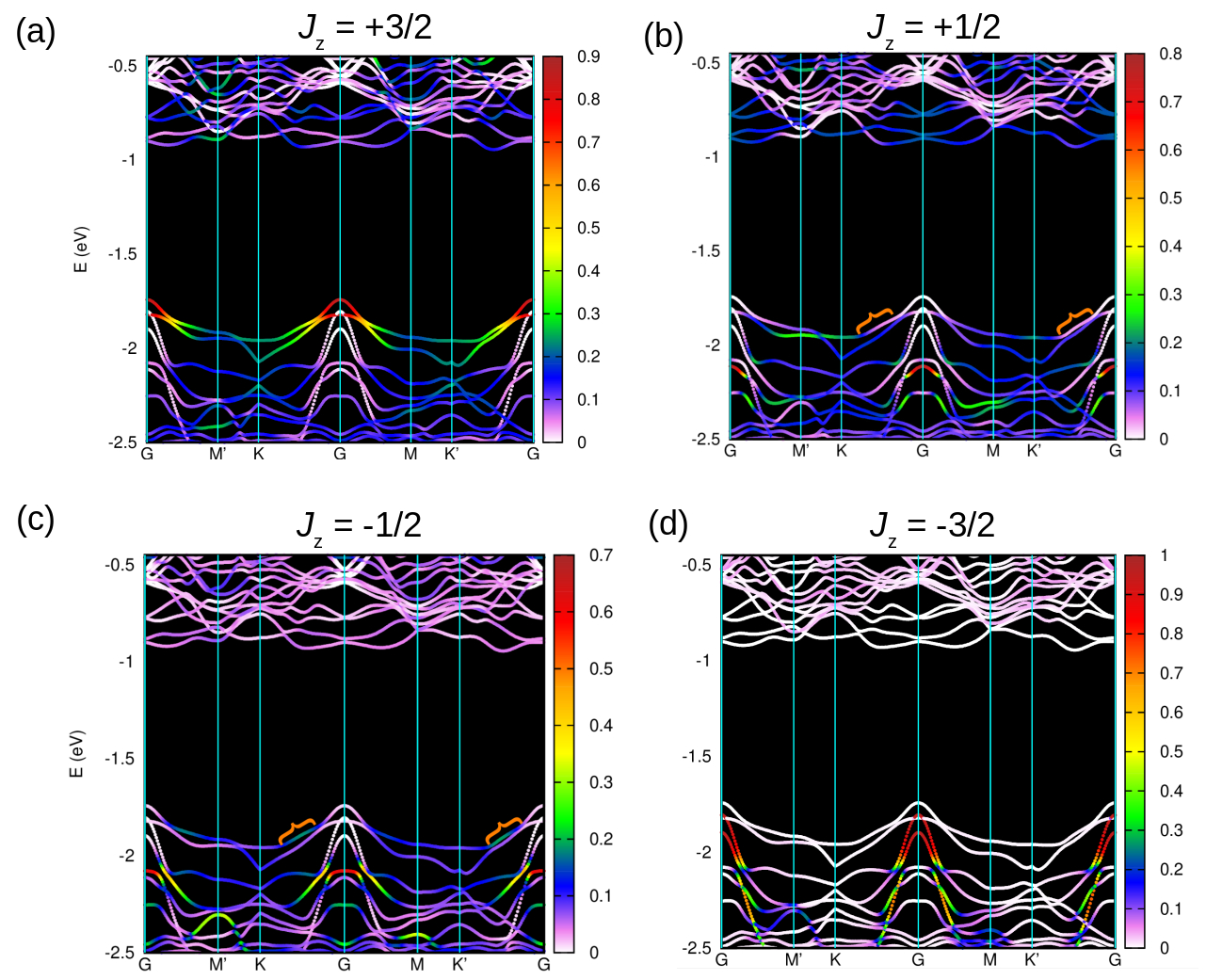}
\caption{Relativistic band-structure plots for the CrI$_{3}$ FM bilayer with projected I-$5p$ atomic spin-orbital characters. The projections are on I-$5p$ $J=3/2$ atomic states with  $J_{z}=+3/2$ (a), $J_z = +1/2$ (b), $J_z = -1/2$ (c), and  $J_z = -3/2$ (d). The energy window includes the highest valence bands and lowest conduction bands. The band-structure regions corresponding to the spin patches along $\Gamma -K$ and  $\Gamma -K'$ are shown by the orange symbols in (b) and (c).}
\label{Jz_0.5_unstrained}
\end{figure*}

In the bilayer band-structure plots, in Fig.~\ref{Jz_0.5_unstrained}, we display  along the bands the I-$5p$ spin-orbital characters projecting on the I-$5p$ $J=3/2$ atomic states with $J_{z}=+3/2$, $+1/2$, $-1/2$, and $-3/2$ in panels (a), (b), (c), and (d), respectively --- the same is reported for the monolayer in the Appendix. Note that the I-$5p$ $J=1/2$ states are located at lower energy than the I-$5p$ $J=3/2$ states seen in Fig.~\ref{Jz_0.5_unstrained}, due to the large I-$5p$ atomic spin-orbit splitting ($0.9$~eV).\cite{SG_2021} 
Near the BZ center, the two upper valence bands of the bilayer have nearly complete I-$5p$, $J=3/2$, $J_{z}=+3/2$ character, as in the monolayer, see Fig.~\ref{Jz_0.5_unstrained}(a). These bands correspond to the highest I-$5p$ $|J=3/2,J_{z}=+3/2\rangle$ states of the monolayers, Fig.~\ref{ml_Jz_0.5}(a), and split at (and near) the $\Gamma$ point into antibonding and bonding states (respectively odd and even under inversion at $\Gamma$) due to the interlayer potential $\Delta V$. 
Comparison of the projected characters in the bilayer and monolayer  along the $\Gamma - K$ and $\Gamma - K'$ directions indicates as main change, due to the interlayer interaction, the inclusion of an I-$5p$ $|J=3/2,J_{z}=+1/2\rangle$ component in the two highest valence states of the bilayer, for $k \approx 0.2-0.3 \cdot 2 \pi/a$, corresponding to the spin patches. At the same time the weight of the I-$5p$ component $|J=3/2,J_{z}=-1/2\rangle $ of these two  states also somewhat increases with respect to that of the monolayer at those $k$ radii. 

It is important to note that inclusion of the $|J=3/2,J_{z}=+1/2\rangle$ component in the highest-valence states of the bilayer, in  Fig.~\ref{Jz_0.5_unstrained}(b), is essential for the occurrence of the in-plane spin components $m_{\parallel}^{l_i}$ on the two layers, as it gives rise to non-vanishing matrix elements  $\langle J, J_{z}|\hat{\sigma}_{\alpha}|J, J'_z\rangle$ for $\alpha = x,y$. This applies to both the Rashba-related and the spin-patch canting effects in the central part of the BZ. 

\begin{figure*} [t]
\includegraphics[width=0.65\textwidth] {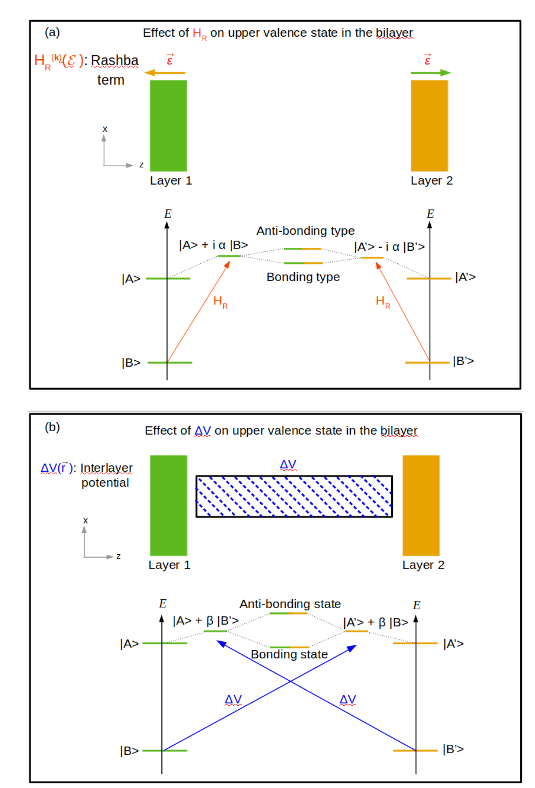}
\caption{Schematics of interlayer Rashba (a) and chiral potential (b) interactions in the FM bilayer and their perturbative effects on the highest valence states of the monolayers, at fixed ${\bm k}$ along the $k_x$ axis (near $\Gamma$).  $|A\rangle$ and $|B\rangle$ ($|A'\rangle$ and $|B'\rangle$) stand for the eigenstates of the monolayer 1 (2). Rashba interaction arises due to the electric field $\mathbfcal{E}$ generated on layer 1 due to layer 2. The Rashba term, $H_{R}$, produces a coupling  between eigenstates of the same monolayer. This coupling induces opposite components in the states of the two layers, and shifts their energy level.  The chiral interlayer potential $\Delta V$ couples eigenstates from different monolayers. This term induces identical components in the states of the two layers, and shifts their energy level. The bonding and antibonding states resulting from (b) [(a)] account for the properties of the $m_x$ [$m_y$] spin-polarization components observed on the two layers for the two upper valence states of the bilayer. }
\label{scheme}
\end{figure*}

Starting thus from the relevant $|J=3/2,J_z=3/2\rangle$  and $|J=3/2,J_z=1/2\rangle$ basis states  of the two isolated monolayers at fixed ${\bm k}$ along the $\Gamma - K'$ direction, and considering as perturbations the local electric field $\mathbfcal{E}^{(i)}$ and the chiral component of the interlayer potential, in the Appendix we determine which intra and interlayer coupling components among the basis states are compatible with the canting behavior found on the two layers for the upper two valence states of the bilayer.  
The type of coupling terms involved in the Rashba and chiral interlayer-potential effect are schematically displayed in Fig.~\ref{scheme}. 
In the following we discuss separately these two types of coupling and canting, and their consequences on the electronic states.

\subsubsection{Spin-canting induced by Rashba term}
\label{sec:Rashba}
 
To describe in general the effect of an electric field $\mathbfcal{E}_z$ on the  $J=3/2$ states of the monolayer, the most appropriate model description, in view of  the large I-$5p$ SOC, is provided by the orbital Rashba Hamiltonian  $H_R^L = - {\bm {d}_{\rm L} \cdot \mathbfcal{E}_{\rm z}} = - \gamma ({\bm k} \times {\mathbfcal{E}_{\rm z}})\cdot {\bm L}$, where ${\bm d}_{\rm L} = \gamma ({\bm L} \times {\bm k})$ is the electronic dipole moment created by the asymmetric charge distribution,\cite{Park_PRL_2011} and $\gamma$ is a proportionality constant. 
Here we are specifically interested in the perturbation effect of the local electric field  $\mathcal{E}_z$ on the $|J=3/2, J_z=+3/2\rangle$ states of the monolayer upper-valence band. In that case, we note that the coupling and the perturbation effect on the upper valence states are actually the same (within a proportionality constant) to those obtained using the phenomenological Rashba spin Hamiltonian  $H_{R}^{{\bm k}}=-{\bm d}_{\rm L}\cdot \mathbfcal{E}_z = - \lambda ({\bm k}\times \mathbfcal{E}_z)\cdot {\boldsymbol \sigma}$, with $\lambda$ taken as a constant for those states. In the following, when using  'Rashba term' we refer in principle to $H_R^L$, but the same is obtained with $H_R$. 
The local electric field $\mathcal{E}_z$ on one monolayer is a measure of the local asymmetry (for that monolayer) in the electrostatic potential induced by the other CrI$_3$ monolayer. 

 Figure~\ref{scheme}(a) shows a schematic representation of the Rashba interaction and its effect in a L\"owdin perturbation approach on the highest valence state of each layer (denoted by $|A\rangle$ and $A'\rangle$ in layer 1 and 2, respectively). 
 The local electric fields on the two layers are opposite, as mentioned earlier. For  ${\bm k}$ around $\Gamma$, the Rashba (onsite) term couples the upper I-$5p$ $|J=3/2,J_z=+3/2\rangle$ state of each layer with the $|J=3/2,J_z=+1/2\rangle$ state of the same layer (non-zero coupling is obtained only for $\Delta J_z = \pm 1$). The interacting bands are denoted in Fig.~\ref{ml_Jz_0.5}.
The $|A\rangle$ state interacts thus, via the Rashba term, with the $|B\rangle$ state from layer 1, and the $|A'\rangle$ state with the $|B'\rangle$ from layer 2. This coupling induces opposite components $i \alpha$~$|B\rangle$ and $-i \alpha$~$|B'\rangle$ in the upper states ($|A\rangle$ and $A'\rangle$) of the two layers, because of the opposite local electric fields on the two layers (with $\alpha$ real, see Appendix \ref{app:Ez}).  It also shifts to higher energy the upper energy level of each layer (with a shift proportional to $|k|^2$). 
  
  The upper two states with equal energy of the two layers  combine then in bonding and antibonding states, 
  as indicated schematically in Fig.~\ref{scheme}(a). The two resulting bonding- and antibonding  states display in-plane spin components which are opposite on the two layers (see Appendix \ref{app:second_state}). This is consistent with the behavior found in Section~\ref{sec:pristine}, i.e., coexistence of opposite spin-polarization components on the two layers in the same non-degenerate state. 
    
   For convenience, in the scheme shown in Fig.~\ref{scheme}(a), we introduced the Rashba coupling first and then
    the coupling induced by the interlayer potential between the $|A\rangle$ and $|A'\rangle$, which gives rise to the bonding and antibonding split states. However, this order is not really relevant in the L\"owdin perturbation approach. The same types of final upper two states and conclusions are obtained with the reversed order in the perturbations, i.e., starting from the coupling giving the bonding-antibonding split states for the A-type and for the B-type bilayer states and then introducing the $H_R$ or $H_L$ coupling. 
   
   Regarding the bonding and antibonding states, we note that the presence of Rashba components in the orbitals  $|A\rangle + i \alpha |B\rangle$ and $|A'\rangle - i \alpha |B'\rangle$ on the two layers does not affect the energy of the resulting bonding and antibonding states to first order in $\alpha$. However, 
   when $k$ increases and the Rashba components become significant, the magnitude of the $|A\rangle$ and $|A'\rangle$ components will tend to decrease due to normalisation. This in turn will increase the energy of the bonding state and decrease that of the antibonding state (to second order in $\alpha$).  This provides a qualitative explanation for the decrease in the splitting between the antibonding and bonding state with increasing $k$ one observes around $\Gamma$ in  Fig.~\ref{Jz_0.5_unstrained}.

   We would like also to emphasize that, although one may use for simplicity the phenomenological Rashba Hamiltonian $H_R$ for the coupling, in order to  understand within a DFT framework what may be the microscopic origin of the canting, it is appropriate to go back to  the orbital Rashba description: $H_L = - {\bm d}_{\rm L} \cdot \mathbfcal{E}_z$ with ${\bm d}_{\rm L} = \gamma  ({\bm L} \times {\bm k})$. The orbital Rashba description highlights as driving force for the canting the alignment of the electric dipole moments with the electric field $\mathbfcal{E}_z$,\cite{Park_PRL_2011} which in turn requires a mixing/coupling between  $|J=3/2, J_z\rangle$ spin-orbital eigenstates (with $\Delta J_z = \pm 1$).

\subsubsection{Spin canting induced by interlayer chiral potential}
\label{sec:Chiral}

The chirality of the layer stacking in the bilayer leads to an interlayer potential $\Delta V$ which includes also a chiral component (as defined in Section~\ref{sec:system}). As pointed out in that section, there are three special directions associated with the structural chirality of the bilayer, which correspond to the directions in which the body diagonals of the Iodine octahedra of the two layers are aligned (see Fig.~\ref{structure}). These three directions are tilted relative  to the $z$ axis with the same tilt angle, and their projections on the $xy$ plane [schematically drawn in Fig.~\ref{structure}(b)] are along the $\Gamma - K'$ and $\Gamma - K$ directions of the BZ.

Figure~\ref{scheme}(b) shows a schematic view, for ${\bm k}$ along the $\Gamma -K'$ direction, of the coupling induced by the chiral potential $\Delta V$ and its effect on the highest valence states of the two layers, in a L\"owdin-type perturbation approach. 
$\Delta V$ couples the $|J=3/2,J_z=+3/2\rangle$ state of layer 1 (hereafter the $|A\rangle$ state) with the $|J=3/2,J_z=+1/2\rangle$ state of layer 2 (the $|B'\rangle$ state).  
Similarly, the $|J=3/2,J_z=+3/2\rangle$ state of layer 2 (the $|A'\rangle$ state) couples with the $|J=3/2,J_z=+1/2\rangle$ state of layer 1 (the $|B\rangle$ state). In fact, $\Delta V$ couples the same types of states as the Rashba term, with the distinction that here the interacting states reside on different layers. 
The interaction occurs between the $p_x$ orbital of one monolayer and the $p_z$ orbital of the other monolayer having the same spin (see Appendix \ref{app:chiral}). 
Since the interlayer potential is real, the coefficient of the coupled $|B'\rangle$ component is real ---in contrast to the coefficient of $|B\rangle$ for the Rashba effect, which is purely imaginary at the same ${\bm k}$. The chiral coupling shifts to higher energy the level of the two upper degenerate states. 

The two degenerate states then combine, as also indicated in Fig.~\ref{scheme}(b),  to give an antibonding and a bonding state (due to the coupling induced by $\Delta V$ between $|A\rangle$ and $|A'\rangle$). The resulting bonding state 
displays radial spin-polarization components which are identical on the two layers, $m_x^{l_1} = m_x^{l_2}$, while the  antibonding state 
exhibits the reversed spin-polarization components on the two layers, $-m_x^{l_1}$, $-m_x^{l_2}$ (see Appendix). This rationalization, based on the off-site couplings $\langle A|\Delta V|B'\rangle=\langle A'|\Delta V|B\rangle$ caused by the chiral potential, accounts for the spin-polarization behavior observed, in Section~\ref{sec:pristine},  for the two upper valence states in the spin patch regions along the $\Gamma - K'$ direction. 

Analogous arguments hold for the spin patches along the other special directions rotated by $\pm 120^0$ from the $k_x$ axis.
It should be noted that an onsite coupling with $\Delta V$ between the $|J=3/2,J_z=+3/2\rangle$ and $|J=3/2, J_z=+1/2\rangle$  states (as is the case of the Rashba-canting effect) would not account for the observed $m_x$ spin-polarization behavior, as the spin-polarization  would be the same for the two upper valence states. 

Similar to the  Rashba-canting discussion, for simplicity the order of the perturbations has been taken with first the non-zero $A$--$B$ coupling and then proceeding with the coupling for the bonding and antibonding states. As in the Rasbha case, however, the final upper two states and conclusions are the same using the reversed order in the perturbations.
It is important to note that the two effects (Rashba and chiral)  co-exist in the upper valence states, because these effects derive from matrix elements between different basis states of the unperturbed system, e.g., $\langle A|H_L|B\rangle$ versus $\langle A|\Delta V|B'\rangle$ corresponding to different off-diagonal elements in the L\"owdin matrix-perturbation approach. The final two upper valence states are the bonding and antibonding  combinations of $|A\rangle + \beta |B'\rangle + i \alpha |B\rangle$ and $|A'\rangle + \beta |B\rangle - i \alpha |B'\rangle$.

Finally, we would like to emphasize that it is only in the main central part of the BZ (including the patch regions) that the behavior of the spin texture is dominated by the contributions of the iodine $5p$ states discussed in this section. For larger $k$ radii, i.e., beyond the patch regions, the weight of the iodine states strongly decreases, while the weight of the Cr $3d$ states becomes substantial.\cite{SG_2021}   Hence, both the Rashba and chiral interlayer-potential effects are expected to vanish at large $k$ radii, as one observes in Fig.~\ref{FM_firstVS_spint} (and indirectly in Fig.~\ref{Pristine_Sz}).

\section{Manipulation of spin texture with vertical compressive strain}
\label{sec:stress}

We have seen that the highest valence band of the FM bilayer is characterized by a rich in-plane spin texture, with vortices and spin patches induced by the interlayer interaction. We also saw that,  although the Rashba effect does not directly contribute to the (space-integrated) in-plane spin texture, it penalizes in energy the bonding states relative to the antibonding states around  $\Gamma$. All these effects are expected to be intensified when applying a vertical compression on the bilayer, possibly even leading to a valence band-edge states with canted spin. In fact, we found previously that new valence band maxima appear outside $\Gamma$, along the $\Gamma - M'$ lines of  the BZ, when the FM bilayer is vertically compressed.\cite{SG_2021} Here we address the corresponding spin texture.

\begin{figure*}[t]
\includegraphics[width=0.95\textwidth] {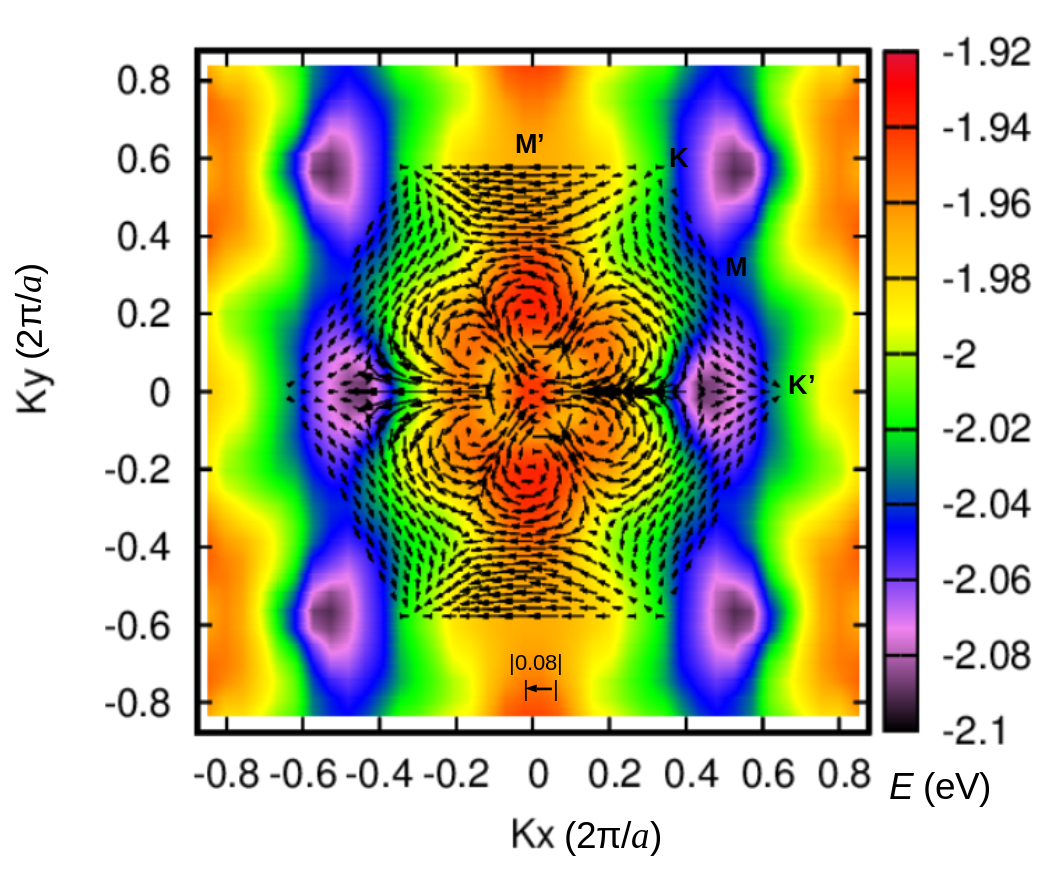}
\caption{\label{spin_texture_strained} Same as in Fig. \ref{FM_firstVS_spint}, but for the FM bilayer with 5\% vertical compression.}
\end{figure*}

Figure~\ref{spin_texture_strained} shows the in-plane spin texture for the upper valence band of the FM bilayer with 5\% compressive strain applied along the vertical direction. The iso-energy map of the band is also displayed in this figure. The compressive strain clearly modifies considerably the in-plane spin-texture pattern and enhances the spin canting, in particular along the $\Gamma - K'$ directions. 
One can observe that the spin patches have been elongated and made much thinner by compression, which also brought the vortices  much closer to the BZ center. This gives rise to a ``flower-like'' pattern in the spin texture around $\Gamma$ in Fig.~\ref{spin_texture_strained}. Strong $S_x$ components are also present at and in the vicinity of the high symmetry $M'$ point. The two valence band maxima located along the $\Gamma - M'$ directions, in Fig ~\ref{spin_texture_strained}, display significant $S_{x}$-spin components. Local valence band maxima outside $\Gamma$ are also present along the $\Gamma - M$ directions and show non-negligible in-plane spins. Hence, the vertical compression produces valence band-edge states with notable in-plane spin canting in the bilayer.

The largest value of $|{\bm S}_{\parallel}({\bm k})|$ appears at $k_{x} = 0.33 \cdot 2\pi/a, k_{y} = 0$, in Fig.~\ref{spin_texture_strained}, and is 0.107. At that ${\bm k}$ point, the amplitude of the in-plane component is $25\%$ of that of the $S_{z}({\bm k})$ component (shown in the SI).  Therefore, the largest ratio of in-plane to out-of-plane spin components is substantially increased upon application of the compressive strain. 
We note that, as in the case of the pristine bilayer, the second highest-valence band displays strong in-plane components along or in the vicinity of the $\Gamma - K'$ and $\Gamma - K$ lines (see SI), with reversed sign relative to the highest valence state. The in-plane spins of the second state are also oriented in the opposite direction, compared to the first state, in the regions of the local valence band maxima of Fig.~\ref{spin_texture_strained}. 

One can notice, in Fig.~\ref{spin_texture_strained}, that the in-plane spin is larger along the special $\Gamma - K'$ and $\Gamma - K$ lines than in any other region in the central part of the BZ. 
As in the case of the unstrained system, we view the vortex features as a consequence of the strong radial spins produced by the chiral potential along these special directions. 
The concentration of vortex features closer to $\Gamma$ is attributed mainly to the intensification of the in-plane chiral spins along those directions at $k \sim 0.2 \cdot 2 \pi /a$ (relative to the BZ-edge spins), which shifts the centers of the vortices towards the BZ center. This shift, however, is also correlated with a peculiar change of the upper valence band in the central part of the BZ (see below).

Figure~\ref{Jz_0.5_strained} shows the projected band-structure plots for the  bilayer with 5\% compression projecting on the iodine $J = 3/2$ atomic states with $J_{z} = +3/2$ (a), $J_{z} = +1/2$ (b), $J_{z} = -1/2$ (c), and $J_{z} = -3/2$ (d). Consistent with the Rashba and chiral coupling discussed in Section~\ref{sec:mechanism}, one observes that the valence-band maxima appearing along $\Gamma - M'$, with canted spin in Fig.~\ref{spin_texture_strained}, display strongly mixed $J_z = +3/2$ and $J_{z}=+1/2$ characters [see Figs.~\ref{Jz_0.5_strained}(a) and (b)]; in fact under compression the $J_{z}=+1/2$ component became the dominant one at the VBM. Similarly, the highest valence states outside $\Gamma$, along the $\Gamma - M$, $\Gamma - K$, and $\Gamma - K'$ direction, with canted spins in Fig.~\ref{spin_texture_strained}, all have strongly mixed  $J_z = +3/2$ and $J_{z}=+1/2$ character. 

One observes, in Fig.~\ref{Jz_0.5_strained}, that under compression the band that corresponds to the VBM outside $\Gamma$ sinks in energy when approaching $\Gamma$, unlike the band that has the highest valence energy at the BZ center. The band which lowers its energy, and has second-highest $J_{z} = +3/2$ character at and near $\Gamma$, corresponds to interlayer bonding states, whereas the other band, with largest $J_{z} = +3/2$ weight at $\Gamma$, corresponds to the antibonding counterparts. Compression increases the bonding-antibonding splitting at $\Gamma$, as expected for layers coming closer to each other. Interestingly, however, the bonding-band states experience outside $\Gamma$ a rapid increase in energy with increasing $k$ radii, which we attribute largely to the Rashba effect, which penalizes the bonding state relative to the antibonding one.

\begin{figure*} [t]
\includegraphics[width=0.98\textwidth] {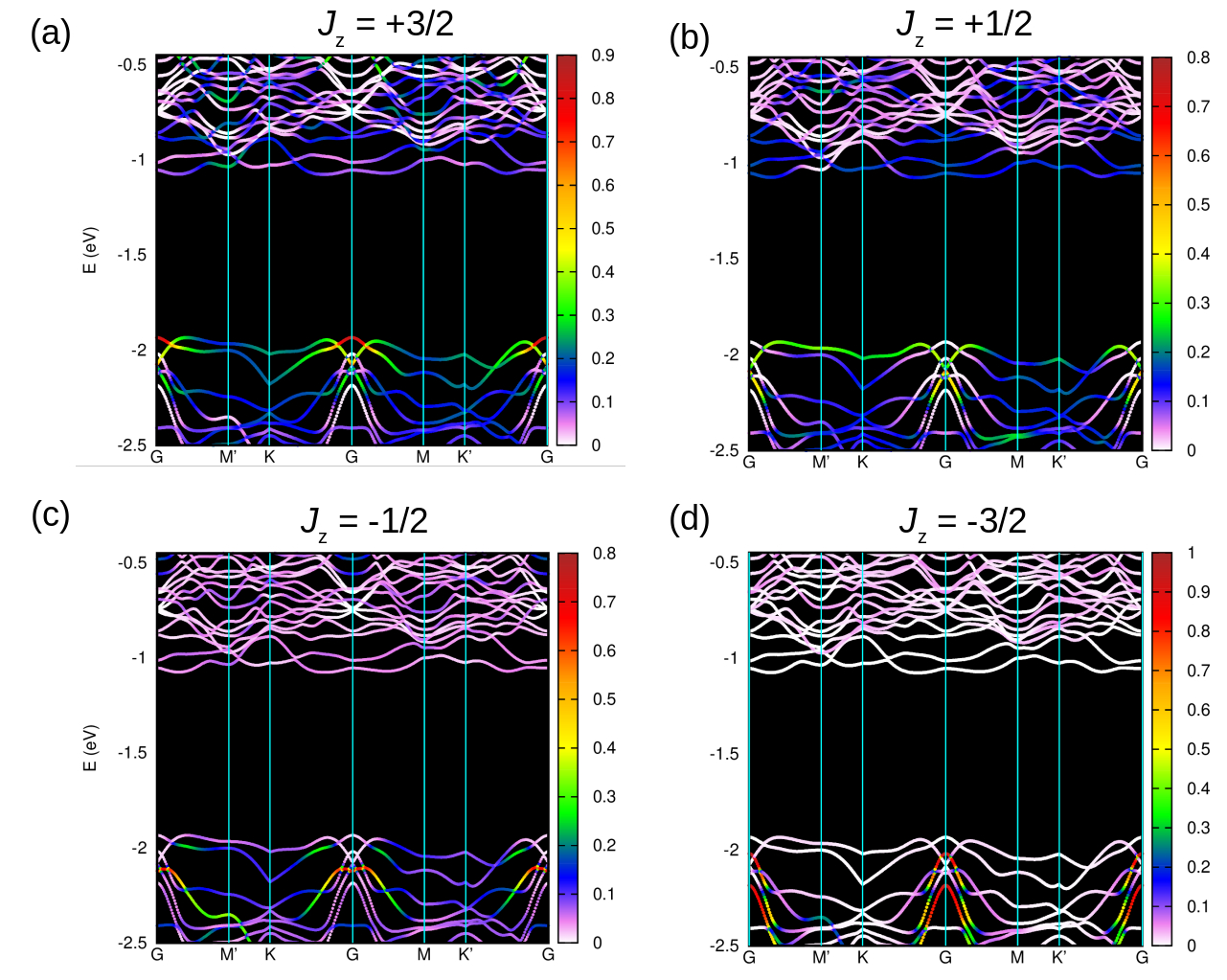}
\caption{\label{Jz_0.5_strained} Same as in Fig.~\ref{Jz_0.5_unstrained}, but for the CrI$_{3}$ FM bilayer with 5\% compressive strain.}

\end{figure*}

    
\section{Conclusions}
\label{sec:conclusion}

In this work we have studied the spin-polarization canting effects and momentum  spin-texture properties in a 2D centrosymmetric ferromagnet with perpendicular magnetization, based on DFT calculations. Using as an example the spin textures of the highest valence bands of the CrI$_3$ FM bilayer, we have shown the existence of two distinct relativistic  spin-polarization canting effects related to the interlayer interaction. 

The first effect derives from the 'chiral' character (lack of reflection symmetry between the two layers) of the interlayer potential, combined with the spin-orbital polarized nature of the highest valence states. This effect is found to induce on both layers of the bilayer the same in-plane spin-polarization canting. In the case of the upper valence band of the CrI$_3$ FM bilayer, this produces radial ``spin-patch'' features in the in-plane-spin texture along specific chiral directions of the BZ. The second effect is a Rashba-related spin-polarization canting, caused by the electric field on each layer due to the presence of the other layer. Interestingly, in the case of a centrosymmetric ferromagnet this effect induces within the same electronic state opposite spin-polarization canting on the two layers. This results in an energy-penalization effect for bonding states relative to antibonding states in a centrosymmetric ferromagnet. 

Furthermore, using the FM bilayer as an example, we have provided some general rules for centrosymmetric systems imposed on the momentum spin texture by magnetic-group symmetry operators. We have shown that symmetry operators that combine time reversal with rotations and reflections in magnetic-space groups of centrosymmetric systems give rise to rules for the spin polarization vector in the BZ which are reversed with respect to those established for non-magnetic groups; namely they impose that spin-polarization vectors must be parallel to mirror planes and perpendicular to rotation axes for such operators. 

In addition, exploiting the above FM chiral and Rashba effects, we have shown for the CrI$_3$ FM bilayer that perpendicular compressive strain can be used to effectively manipulate the spin texture. In particular, our {\it ab initio} calculations for the CrI$_3$ FM bilayer show that vertical compression induces valence-band-edge states with canted spin polarization. 

The  chiral and Rashba spin-canting FM effects disclosed in the present study are expected to be observed also in other 2D centrosymmetric FM bilayers with SOC and, respectively, chiral layer stacking and semiconductor character.  Furthermore, the present rules on spin textures for centrosymmetry systems imposed by magnetic-group operators complement those established for the non-magnetic materials, and are expected to be helpful guidelines for the shapes of spin textures in BZ's of such magnetic materials in general. 

\setcounter{figure}{0}

\setcounter{section}{0}    

\renewcommand{\thesection}{A}   
\renewcommand{\thesubsection}{A\arabic{subsection}}

\section{Appendix}

\subsection{CrI$_{3}$ FM monolayer: spin texture and projected band structure}

\begin{figure*}[t]
\includegraphics[width=0.7\textwidth] {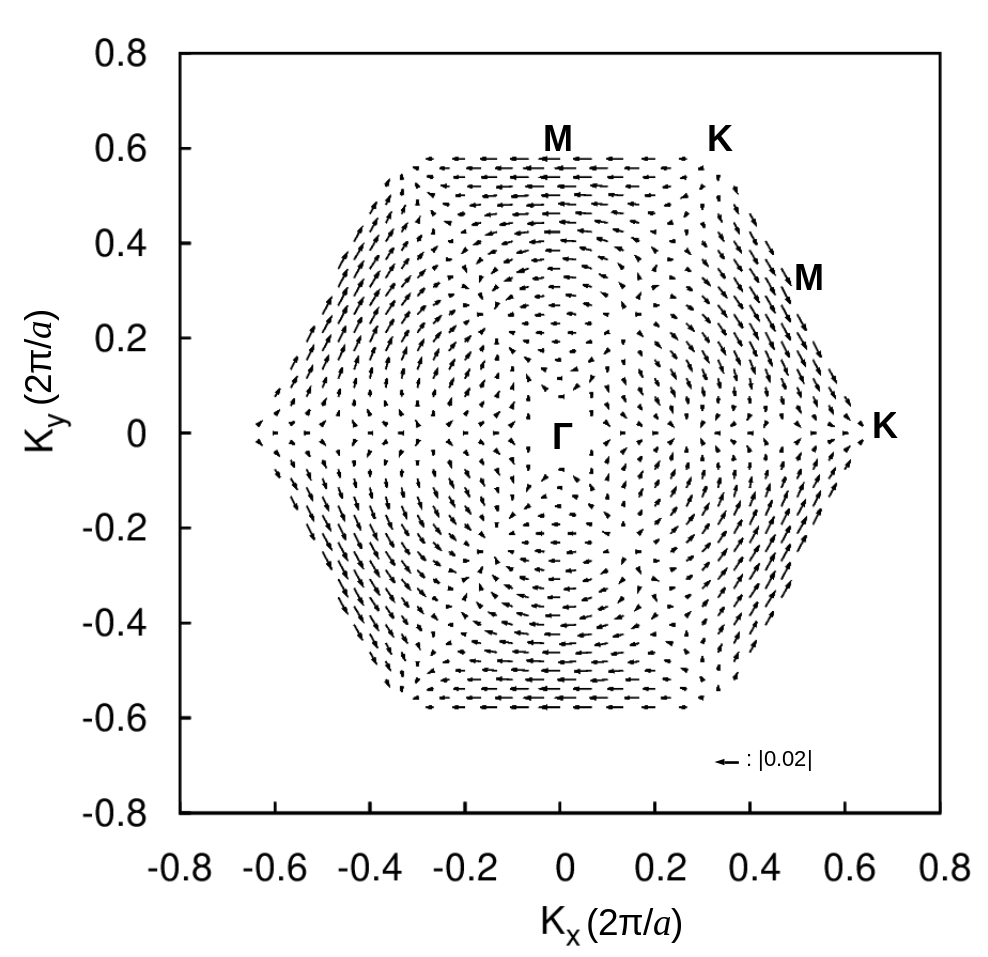}
\renewcommand{\thefigure}{A\arabic{figure}}
        \caption{\label{ml_st} In-plane spin texture for the highest valence band of the CrI$_{3}$ FM monolayer in the 2D BZ.} %
\end{figure*}

Figure~\ref{ml_st} shows the in-plane spin texture for the highest valence band of the CrI$_{3}$ FM monolayer. The in-plane spin texture is seen to be non-negligible essentially only near the BZ edges. The highest magnitude of the in-plane spin,  ${\bm S_{\parallel}}$ appear in close vicinity of the $M$ and $M'$ points and is 0.021. 

The spin-texture pattern observed along the contour of the BZ, in Fig.~\ref{ml_st}, with spins parallel to the BZ edges, is imposed by the symmetry rules of  Section~\ref{sec:sym}, for the magnetic group D$_{3d}$[S$_6$] of the monolayer. This group includes, in particular, the symmetry operators $R_{y}(180^0) \cdot T$, where time reversal is combined with the two-fold rotations about the $y$ and equivalent directions (along the $\Gamma - M$, and $M' - \Gamma$ directions).  
The corresponding symmetry rules impose that for ${\bm k} \parallel \Gamma -M^{(')}$,  ${\bm S}({\bm k}) \perp \Gamma-M^{(')}$. 
Furthermore, the monolayer group also includes the symmetry operators $R_{z}(\pm 60^0)\cdot M_{xy}$. The corresponding symmetry rules impose that ${\bm S_{\parallel}}(M') = -R_{z}(60^0){\bm S_{\parallel}}(M)$. 

We note that the monolayer magnetic group also contains the  symmetry operators $M_d \cdot T$, where time reversal is combined with the reflections about vertical mirror planes parallel to the $x$ axis and equivalent directions ($\Gamma - K'$ and $K - \Gamma$). The corresponding symmetry rules imply that for ${\bm k} \parallel \Gamma -K^{(')}$,  ${\bm S_{\parallel}}({\bm k}) \parallel \Gamma-K^{(')}$. 

\begin{figure*}[t]
\includegraphics[width=0.95\textwidth]{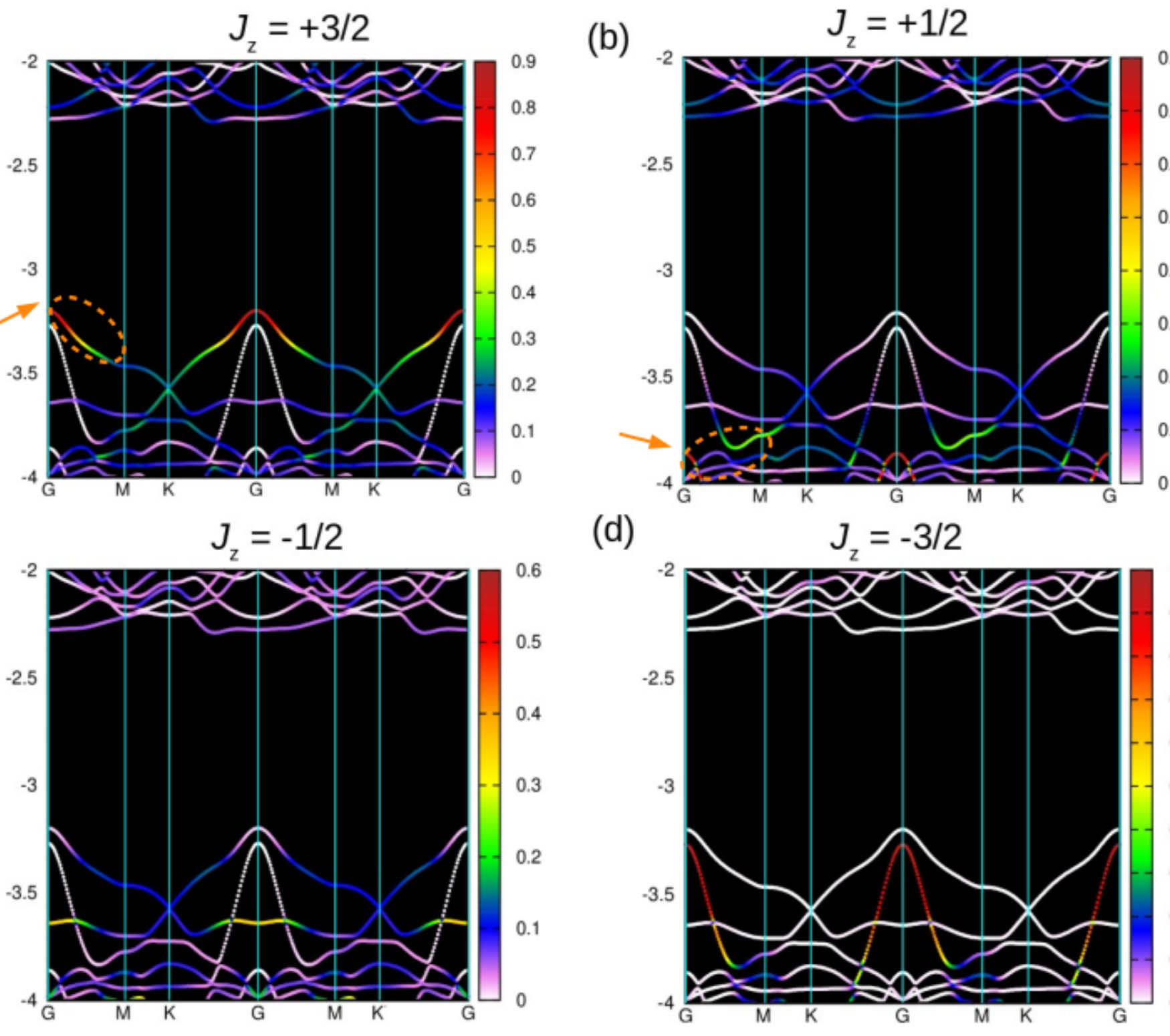}
\renewcommand{\thefigure}{A\arabic{figure}}
        \caption{\label{ml_Jz_0.5} The same as in Fig.~\ref{Jz_0.5_unstrained}, but for the CrI$_3$ monolayer. The bands interacting through the Rashba term are denoted by the orange ellipses in (a) and (b)  - {\it e.g.} the first and fourth band at $\Gamma$. }   %
\end{figure*}

Figure~\ref{ml_Jz_0.5} shows the projected band-structure plots for the CrI$_3$ FM monolayer projecting on the iodine $J = 3/2$ atomic states with $J_{z} = +3/2$ (a), $J_{z} = +1/2$ (b), $J_{z} = -1/2$ (c), and $J_{z} = -3/2$ (d). 
The CrI$_3$ monolayer is characterized by an upper valence band which is isolated in energy in the central part of the BZ. The corresponding states are predominantly I-$5p$ spin-orbital states, with total momentum $J=3/2$ and component $J_z=3/2$. 
Apart from the main $|J=3/2,J_z=3/2\rangle$  component, the highest valence band also includes a small, but non-negligible, I-$5p$ $|J=3/2,J_z=-1/2\rangle$ component for $k$ radii $0.2-0.3 \cdot 2 \pi/a$. The states of the upper valence band have vanishing in-plane  ${\bm S}_{\parallel}({\bm k})$ in the central part of the BZ, as a result of the rule: $\langle J, J_{z}|\hat{\sigma}_{\alpha}|J, J'_{z}\rangle \neq 0$, $\alpha = x,y$, only when $\Delta J_z = \pm 1$. We also denote, in Fig.~\ref{ml_Jz_0.5}, the bands which interact through the Rashba term, see Section~\ref{sec:Rashba}. We note that the monolayer $J_z=+1/2$ states with $p_z$ component move higher in energy upon formation of the bilayer.\cite{SG_2021}

\subsection{CrI$_{3}$ FM bilayer: spin texture of second-highest valence band}  \label{app:second_state}

\begin{figure*}[t]
\includegraphics[width=0.7\textwidth]{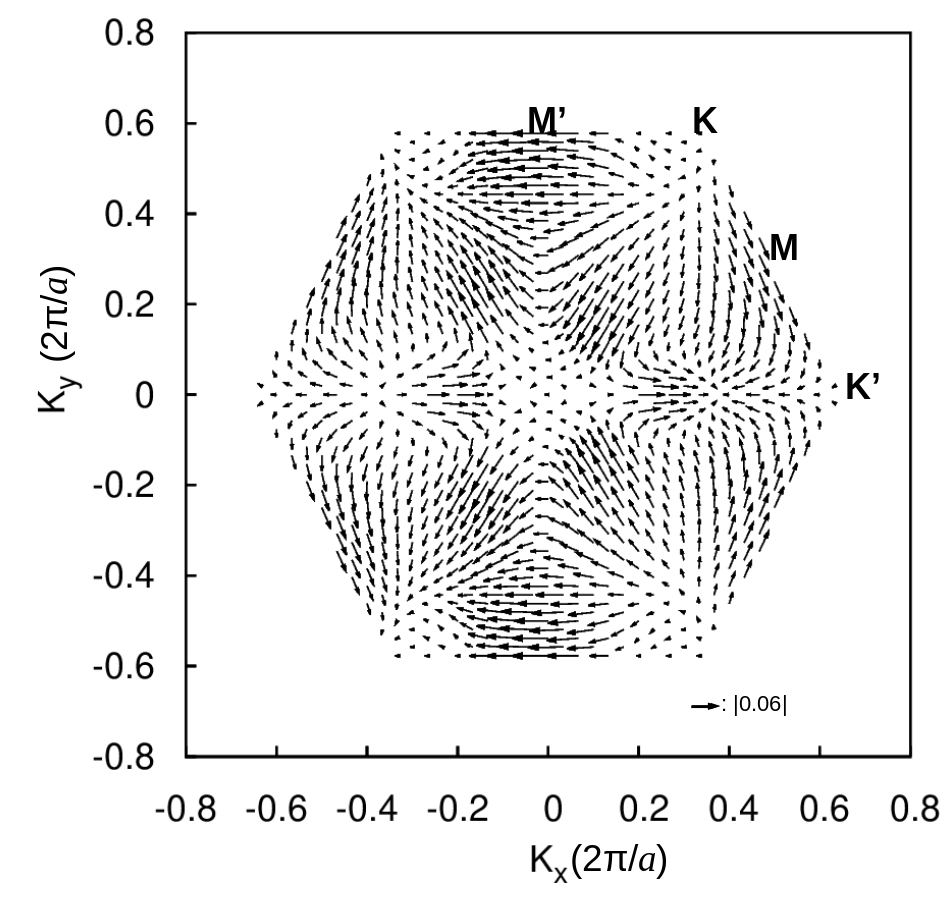}
\renewcommand{\thefigure}{A\arabic{figure}}
        \caption{\label{VS_1_bl} In-plane spin texture for the second highest valence band of the CrI$_{3}$ FM bilayer.} %
\end{figure*}

Figure~\ref{VS_1_bl} shows the in-plane spin texture plot for the second highest valence band of the pristine FM bilayer. Comparing to the spin texture plot of the highest valence band, in Fig.~\ref{VS_1_bl}, one sees that the in-plane spin polarization components of the first and second highest valence states are opposite in the regions of the six spin patches. 

\subsection{Implications of magnetic group symmetries on spin textures and related quantities}    
\label{app:symmetries}

 To establish in general the symmetry properties of the spin texture and  spin-polarization density in momentum space,  considering the magnetic group of a given magnetic system,  we distinguish between operators involving time reversal and those which do not involve time reversal. We consider non-degenerate states (as normally is the case in a FM system) at ${\bm k}$ points around $\Gamma$, i.e., with ${\bm k}$ relative to the BZ center. For degenerate states, the same result on symmetries applies to the generalized quantity obtained by summing over the degenerate states. 

For an operator that does not involve time reversal, i.e., a spatial symmetry operation, $\{R|{\bm f}\}$, where $R$ is a point-group operation and ${\bm f}$ a fractional translation, the fact that $\{R|{\bm f}\}$ commutes with H$_{KS}$ implies for the energy eigenvalues: E$_n(R{\bm k})$ = E$_n({\bm k})$. Furthermore, for the non-degenerate eigenstates at ${\bm k}$, it implies for the corresponding  probability densities:    $|\psi_{n,R{\bm k}}|^2({\bm r}) = |\psi_{n,{\bm k}}|^2(R^{-1}{\bm r}-R^{-1}{\bm f})$, for the spin-polarization densities: ${\bm m}^{(n, R{\bm k})}({\bm r}) = \{R|{\bm f}\} {\bm m}^{(n,{\bm k})}({\bm r})$, and 
for the  spin expectation values: ${\bm S}(n, R{\bm k}) = R{\bm S}(n, {\bm k})$. 

Time reversal is acting on spinors as:\cite{Bassani_1975} $T \psi_{(n,{\bm k})}({\bm r}) = i \hat{\sigma}_y \psi^{*}_{n,{\bm k}}({\bm r})$. Taking this into account, and considering then a spatial operator $\{R'|{\bm f'}\} T$ of the magnetic point group, with $R'$ a spatial point-group symmetry operation and ${\bm f'}$ a fractional translation, the  commutativity of $\{R'|{\bm f'}\} T$ with H$_{KS}$ implies for the energy eigenvalues: E$_n(-R'{\bm k})$ = E$_n({\bm k})$. For the non-degenerate eigenstates at ${\bm k}$, it  implies for the associated  probability densities:   $|\psi_{n,-R'{\bm k}}|^2({\bm r}) = |\psi_{n,{\bm k}}|^2(R'^{-1}{\bm r}-R'^{-1}{\bm f'})$, 
for the spin-polarization densities:  ${\bm m}^{(n, -R'{\bm k})}({\bm r}) =   \{R'|{\bm f'}\}T$~${\bm m}^{(n,{\bm k})}({\bm r})$, and for the spin expectation values: ${\bm S}(n, -R'{\bm k}) = R'T$~${\bm S}(n, {\bm k})$.

The spin expectation values, ${\bm S}$, and the spin-polarization density,  ${\bm m}$, are axial vectors (vectors with a virtual current loop), and transform accordingly under time reversal (i.e., $T{\bm S} = -{\bm S}$ and $T{\bm m} = -{\bm m}$) and under spatial operations $R$, $R'$.\cite{Chapon_2012} Using this and the above relations for ${\bm S}$ and ${\bm m}$, with the operators of the magnetic group of the system under consideration, yields the correspondence rules for ${\bm S}$ and ${\bm m}$ at ${\bm k}$ points related by the symmetry operators. In the specific case of the C$_{2h}$[C$_i$] magnetic point group, this yields the correspondence rules given in Table~\ref{Symm_oper}.

\subsection{\label{perturbation1}{Effect of $\mathbfcal{E}_z$ on the monolayer highest valence states}}
\label{app:Ez}

We are interested in the perturbation effect of the Rashba term $H_R^L$ (or $H_R$) on the states of the monolayer highest-valence band in the central part of the BZ, i.e., with $J_z = +3/2$. We work at fixed ${\bm k}$ and consider ${\bm k}$ along the $k_x$ axis. 
The four highest valence states of the monolayer at and near $\Gamma$ correspond essentially to the four different I-$5p$ $J=3/2$ atomic spin-orbital states on the Iodine atoms with $J_z = \pm 3/2$ and $\pm 1/2$ (see Fig. \ref{ml_Jz_0.5}), where the $z$ axis is along the spins of the Cr atoms (see Fig.\ref{structure}). 
The Rashba term $H_R^L$ (or $H_R$) couples only onsite states with $\Delta J_z = \pm 1$. Hence, the highest valence state with  $J_z = +3/2$  couples via $H_R^L$ only to the $J_z=+1/2$ state of the same monolayer. 
In Fig.~\ref{scheme}(a), for layer 1, the $J_z = +3/2$ state is $|A\rangle$ and the $J_z=+1/2$ state is $|B\rangle$, and the Rashba coupling is given by $\langle A|H_R^L| B\rangle=\langle B|H_R^L|A\rangle^* \neq 0$. 

As the difference between the two energy levels, $E_A - E_B$, is significantly larger than the magnitude of their coupling  $|\langle A|H_R^L| B\rangle|$, the L\"owdin approach is used to include the correction induced by $\langle B|H_R^L| A\rangle$ to the diagonal matrix element of the Hamiltonian of the highest valence state $|A\rangle$ on layer 1:
   $ \Delta E = \frac{|\langle A|H_\mathrm{R}^L|B\rangle|^{2}}{E-E_{B}}$,
where $E \approx E_{A}$. The same holds for $|A'\rangle$ and $|B'\rangle$ on layer 2. The Rashba term $H_R^L$ for layer 2, however, has the opposite sign relative to layer 1, due to the opposite local electric field. Taking this into account and considering also the coupling $\langle A|\Delta V|A'\rangle$ between states of the two layers, the L\"owdin-corrected bonding and antibonding type states are proportional to: 
$|A\rangle + \chi |B\rangle \pm (|A'\rangle - \chi |B'\rangle)$, with $\chi = \frac{\langle B|H_\mathrm{R}^L|A\rangle}{E - E_B}$. Using for $|A\rangle$ and $|B\rangle$, the spin-orbital states $|J_z=3/2\rangle$ and $|J_z=1/2\rangle$ given in the SI, $\chi$ is found to be purely imaginary ($\chi = i \alpha$, with $\alpha$ real), and involves coupling between $p_x$ orbital on one layer and $p_z$ orbital on the other layer. For ${\bm k}$ along the $k_x$ axis  
the corresponding in-plane spin-polarization components for the `'bonding'' and ``antibonding'' states on the two layers are along the $y$ axis, and are opposite on the two layers (and same for the two states).

\subsubsection{Effect of the interlayer chiral potential on the highest valence states}
\label{app:chiral}

We have seen in Section~\ref{sec:sym} that the interlayer chiral potential $\Delta V({\bm r})$ is responsible for the in-plane spin texture. Furthermore, we have seen, in Section~\ref{sec:pristine}, that for the upper valence states the in-plane spin texture requires coupling between the $J_z=3/2$ and $J_z=1/2$ states. Here we examine which type of coupling between such states (intralayer versus intralayer) can explain the main behavior of the spin-polarization density observed, in Section~\ref{sec:pristine}, for the two upper valence states with ${\bm k}$ along the $k_x$ axis (in the spin patch). 

We consider first the interlayer coupling between $|J_{z}=+3/2\rangle$ and $|J_{z}=+1/2\rangle$ states induced by $\Delta V$, i.e., the $\langle A|\Delta V|B'\rangle=\langle B'|\Delta V|A\rangle^*$ and $\langle A'|\Delta V |B\rangle=\langle B|\Delta V|A'\rangle^*$ matrix elements. As in the Rashba case, the difference between the energy levels, $E_A - E_B$, is large compared to the magnitude of these matrix elements, 
whose effect on the upper valence states is treated within the L\"owdin perturbation approach. The corresponding correction to the diagonal matrix element for the highest valence state $|A\rangle$ on layer 1 is given by:  $\Delta E = \frac{|\langle B'|\Delta V|A\rangle|^{2}}{E-E_{B'}}$, with $E \approx E_A$. The same holds for layer 2. As $\Delta V$ is inversion symmetric, one has $\langle A|\Delta V|B'\rangle=\langle A'|\Delta V|B\rangle$. Taking this into account and considering also the interlayer coupling $\langle A| \Delta V|A'\rangle$, the  corresponding L\"owdin-corrected bonding and antibonding type states are proportional to: $|A\rangle+ \beta |B'\rangle \pm (|A'\rangle+\beta|B\rangle)$, with $\beta = \frac{\langle B'|\Delta V|A\rangle}{E-E_{B'}}$. 
Using for $|A\rangle$ and $|B'\rangle$, the spin-orbital states $|J_z=3/2\rangle$ on layer 1 and $|J_z=1/2\rangle$ on layer 2 of the SI for ${\bm k}$ along the $k_x$ axis,  and $E \approx E_A$, the corresponding bonding and antibonding states yield in-plane spin-polarization components, $m_x$, which are the same on the two layers, and opposite for the two states. This is consistent with the behavior observed for the two highest valence states of the bilayer in  Section~\ref{sec:pristine}. 

Considering, instead, as main $A$--$B$ type of coupling the intralayer coupling $\langle A|\Delta V|B\rangle$ and $\langle A'|\Delta V| B'\rangle$, the corresponding L\"owdin-corrected bonding and antibonding type states would be proportional to: $|A\rangle+ \beta' |B\rangle \pm (|A'\rangle+\beta'|B'\rangle)$, with $\beta' = \frac{\langle B|\Delta V|A\rangle}{E-E_B}$. 
These states yield $m_x$ spin-polarization components which are the same for the two states, and would therefore not account for the spin-polarization behavior for the upper two valence states obtained from the {\it ab initio} calculations. Hence, it is the coupling induced by  $\Delta V$ between $J_z=3/2$ and $J_z=1/2$ states of different layers which controls the trends observed in the in-plane spin texture.

\bibliography{CrI3_aps}



\end{document}